\documentclass[a4paper,11pt]{article}
\usepackage{jheppub}

\usepackage[utf8]{inputenc}
\usepackage{graphicx,cancel,float}
\usepackage{amssymb}
\usepackage{textcomp}
\usepackage{amsmath}
\usepackage{tabularx}
\usepackage{bm}
\usepackage{times}
\usepackage{color}
\usepackage{slashed}
\usepackage{multirow}
\usepackage{verbatim}
\usepackage{epsfig,epstopdf}

\allowdisplaybreaks[4]

\def\lsim{\mathrel{\raise.3ex\hbox{$<$\kern-.75em\lower1ex\hbox{$\sim$}}}}
\def\gsim{\mathrel{\raise.3ex\hbox{$>$\kern-.75em\lower1ex\hbox{$\sim$}}}}

\newcommand{\calO}{{\mathcal{O}}}

\definecolor{orange}{rgb}{1,0.5,0}

\usepackage[table]{xcolor}
\usepackage{colortbl}
\definecolor{mygray}{gray}{.95}

\preprint{}
\subheader{\it \today}

\title{Constraints on the charged currents in general neutrino interactions with sterile neutrinos}

\author[a]{Tong Li}
\emailAdd{litong@nankai.edu.cn}
\affiliation[a]{School of Physics, Nankai University, Tianjin 300071, China}
\author[b]{Xiao-Dong Ma}
\emailAdd{maxid@phys.ntu.edu.tw}
\affiliation[b]{Department of Physics, National Taiwan University, Taipei 10617, Taiwan}
\author[c]{Michael A. Schmidt}
\emailAdd{m.schmidt@unsw.edu.au}
\affiliation[c]{School of Physics, The University of New South Wales, Sydney, New South Wales 2052, Australia}

\abstract
{In this work we investigate the implication of low-energy precision measurements on the quark-lepton charged currents in general neutrino interactions with sterile neutrinos in effective field theories. The physics in low-energy measurements is described by the low-energy effective field theory extended with sterile neutrinos (LNEFT) defined below the electroweak scale. We also take into account renormalization group running and match the LNEFT onto the Standard Model (SM) effective field theory with
sterile neutrinos (SMNEFT) to constrain new physics (NP) above the electroweak scale. The most sensitive low-energy probes are from leptonic decays of pseudoscalar mesons and hadronic tau lepton decays in terms of precise decay branching fractions, the lepton flavor universality and the Cabibbo-Kobayashi-Maskawa (CKM) unitarity. We also consider other constraints including nuclear beta decay. The constraints on charged current operators are generally stronger than the ones for quark-neutrino neutral current operators. We find that the most stringent bounds on the NP scale of lepton-number-conserving and lepton-number-violating operators in SMNEFT are 74 (110) TeV and 9.8 (13) TeV, respectively, for the operators with down (strange) quark.
}

\arxivnumber{arXiv:2007.15408}

\begin{document}

\maketitle
\setcounter{page}{2}

\newpage

\section{Introduction}
\label{sec:Intro}

The absence of any signal for new physics (NP) beyond the Standard Model (BSM) at the Large Hadron Collider (LHC) has sparked a renaissance in the search for the BSM physics at low-energy experiments. The hints for BSM physics, however, may hide in deviations from the SM in low-energy precision measurements. The precision era in neutrino physics sheds light on possible new dynamical degrees of freedom such as right-handed (RH) neutrinos and non-standard neutrino interactions. The results of low-energy precision measurements can guide our direct search for NP in the neutrino sector at future high-energy colliders.

The effective field theory (EFT) below the electroweak (EW) scale can well describe the physics in low-energy measurements and serves as a model-independent way to study the implications for neutrino physics. The low-energy effective field theory (LEFT) is an EFT defined below the electroweak scale $\Lambda_{\rm EW}\sim 10^2$ GeV~\cite{Jenkins:2017jig}.
The LEFT respects the unbroken gauge symmetries $SU(3)_c\times
U(1)_{\rm em}$ after integrating out the Higgs boson $h$, weak gauge bosons $W, Z$ and the top quark $t$ in the SM. The LEFT extended by right-handed neutrinos $N$ is correspondingly named as LNEFT~\cite{Chala:2020vqp,Li:2020lba}. In Ref.~\cite{Li:2020lba} we constructed the complete and independent operator basis for the LNEFT up to dimension-6~\footnote{An independent set of operators at dim-6 in LNEFT was given in Ref.~\cite{Chala:2020vqp}.} and matched the quark-neutrino neutral current interactions in the LNEFT to the SM effective field theory extended by RH neutrinos $N$ (SMNEFT)~\cite{delAguila:2008ir} at the electroweak scale. The SMNEFT respects the SM gauge group $SU(3)_c\times SU(2)_L\times U(1)_Y$ and describes the physics above the electroweak scale up to the NP scale. The full classification of the SMNEFT operators up to dim-7 has been done in Refs.~\cite{Grzadkowski:2010es,Lehman:2014jma,Bhattacharya:2015vja,Liao:2016hru,Liao:2016qyd}.
\footnote{We would like to briefly mention that in recent years there has been progress in automatizing the construction of operator bases. See e.g.~Ref.~\cite{Banerjee:2020bym} for the recently developed Mathematica package GrIP and references therein for other approaches.}
By studying the implication of low-energy measurements for the neutral currents in LNEFT, we found that the most stringent bound on the NP scale in SMNEFT is 1.5 TeV for the neutrino-quark operators~\cite{Li:2020lba}. We note that, after the electroweak symmetry breaking, some SMNEFT operators yielding the neutrino-quark neutral currents can also induce charged current operators in the LNEFT. The charged current operators are made of a neutrino, a charged lepton, an up-type quark and a down-type quark.
More intriguingly, the relevant SMNEFT operators may be subject to more stringent constraints due to the presence of charged leptons.

In this paper we investigate the implication of low-energy measurements on the charged currents in LNEFT with both lepton-number-conserving (LNC) and lepton-number-violating (LNV) operators. The most sensitive low-energy probes arise from the weak leptonic decays of pseudoscalar mesons and hadronic tau lepton decays in terms of precise decay branching ratios (BRs), the lepton flavor universality (LFU) in pseudoscalar meson and tau lepton decays and the Cabibbo-Kobayashi-Maskawa (CKM) unitarity. We also consider the constraint on charged currents in LNEFT from nuclear beta decay and predict new contributions to weak decays of light vector mesons.
Since the high scale NP effect is usually parameterized in the SMNEFT, we then take into account renormalization group (RG) running effect from the experimental scale to the EW scale and match the LNEFT onto the SMNEFT to constrain the relevant NP scale.

The paper is outlined as follows. In Sec.~\ref{sec:GNI}, we describe the
general charged current operators in the LNEFT basis. The LNEFT operators are
then matched to the SMNEFT. We give the analytical expressions for the low-energy constraints from the decay branching fractions, the LFU in pseudoscalar meson and tau lepton decays, the CKM unitarity and nuclear beta decay in Sec.~\ref{sec:Const}. The numerical results are given in Sec.~\ref{sec:Num}. We also discuss other relevant constraints and predict modification of weak decay rates of vector mesons in Sec.~\ref{sec:Other}. Our conclusions are drawn in Sec.~\ref{sec:Con}. Some
calculational details are collected in the appendices.

\section{General neutrino interactions with RH neutrinos}
\label{sec:GNI}

The main focus of this work is on low-energy precision measurements of charged current processes involving neutrinos from the perspective of effective field theory. We in particular consider leptonic decays of pseudoscalar mesons, hadronic tau lepton decays, the LFU in pseudoscalar meson and tau lepton decays, the CKM unitarity and $\beta$ decays.
All those observables are measured at the sub-GeV scale but generated by the $W$ boson at the EW scale and/or the possible heavy NP beyond the EW scale. Thus, it is suitable to work in the framework of LNEFT defined below the electroweak scale $\Lambda_\textrm{EW}$. Its dynamical degrees of freedom include
the five quarks $(u,d,s,c,b)$, all charged leptons $(e,\mu,\tau)$ and neutrinos $(\nu_{e}, \nu_\mu, \nu_\tau)$ in the SM and an arbitrary number of BSM RH neutrinos $N$. The power counting of LNEFT is determined by both the NP scale $\Lambda_{\rm NP}$ and the electroweak scale $\Lambda_{\rm EW}$. The LNEFT consists of dim-3 fermion mass terms, dim-4 kinetic terms and higher dimensional operators $\calO_{i,L}^{(d)}(d\geq 5)$ (dim-$d$) built out of those light fields and satisfies the $SU(3)_c\times U(1)_{\rm em}$ gauge symmetry. The LNEFT Lagrangian is
\begin{align}
\mathcal{L}_{\rm LNEFT}=\mathcal{L}_{\rm d\leq 4}+ \sum_i \sum_{d\geq 5} C_{i,L}^{(d)} \calO_{i,L}^{(d)}\;,
\end{align}
where $C_{i,L}^{(d)}$ is the Wilson coefficient (WC) of operator $\calO_{i,L}^{(d)}$. Generally, the Wilson coefficient $C_{i,L}^{(d)}$ scales as $\Lambda_{\rm EW}^{n+4-d}/\Lambda_{\rm NP}^{n}$ with integer $n\geq 0$. In Appendix~\ref{sec:LNEFTbasis} we summarize the dim-6 operator basis involving RH neutrinos $N$ in the LNEFT~\cite{Li:2020lba}.

We assume the LNEFT is a low-energy version of the SMNEFT which is defined above the electroweak scale.
In the SMNEFT, the renormalizable SM Lagrangian is extended by the RH neutrino sector and a tower of higher dimensional effective operators $\calO_i^{(d)}$ with increasing canonical dimension $d\geq 5$.
The importance of these operators is measured by the Wilson coefficients $C_i^{(d)}$ with decreasing relevance
\begin{eqnarray}
\mathcal{L}_{\rm SMNEFT}= \mathcal{L}_{\rm SM+N} + \sum_i \sum_{d\geq 5} C_i^{(d)} \calO_i^{(d)} \; ,
\end{eqnarray}
where $\mathcal{L}_{\rm SM+N}$ is the renormalizable SM Lagrangian extended by RH neutrinos $N$.
The unknown Wilson coefficient $C_i^{(d)}$ encodes the heavy NP contribution and is associated with an effective NP scale via $\Lambda_{\rm NP}=(C_i^{(d)})^{1/(4-d)}$. For a given NP model, after integrating out the new heavy states, it can be determined as the function of the parameters in the NP model through matching and renormalization group running. In Appendix~\ref{sec:SMNEFTbasis} we collect the relevant SMNEFT operators used in our analysis for the generic neutrino interactions.

\subsection{General lepton-quark charged current operators in LNEFT basis}
\label{sec:LEFT}
Denoting the SM left-handed neutrinos as $\nu$ and the right-handed sterile neutrinos as $N$, the dim-6 quark-lepton charged current operators with lepton number conservation are~\cite{Jenkins:2017jig,Chala:2020vqp,Li:2020lba}
\begin{align}
\calO_{ud\ell\nu1}^V=&(\overline{u_L}\gamma^\mu d_L)(\overline{\ell_L}\gamma_\mu \nu)\;,
&
\calO_{ud\ell\nu2}^V=&(\overline{u_R}\gamma^\mu d_R)(\overline{\ell_L}\gamma_\mu \nu)\;,
\nonumber\\
\calO_{ud\ell\nu1}^S=&(\overline{u_R}d_L)(\overline{\ell_R}\nu)\;, \
&
\calO_{ud\ell\nu2}^S=&(\overline{u_L}d_R)(\overline{\ell_R}\nu)\;,
\nonumber\\
\calO_{ud\ell\nu}^T=&(\overline{u_R}\sigma^{\mu\nu}d_L)(\overline{\ell_R}\sigma_{\mu\nu}\nu)\;,
\label{ope:udlNu}
\\
\calO_{ud\ell N1}^V=&(\overline{u_L}\gamma^\mu d_L)(\overline{\ell_R}\gamma_\mu N)\;,
&
\calO_{ud\ell N2}^V=&(\overline{u_R}\gamma^\mu d_R)(\overline{\ell_R}\gamma_\mu N)\;,
\nonumber\\
\calO_{ud\ell N1}^S=&(\overline{u_L}d_R)(\overline{\ell_L}N)\;,
&
\calO_{ud\ell N2}^S=&(\overline{u_R}d_L)(\overline{\ell_L}N)\;,
\nonumber\\
\calO_{ud\ell N}^T=&(\overline{u_L}\sigma^{\mu\nu}d_R)(\overline{\ell_L}\sigma_{\mu\nu}N)\;,
\label{ope:udlN}
\end{align}
together with their hermitian conjugates.
Here $u_i$ and $d_i$ stand for up-type quarks $(u,c)$ or down-type quarks $(d,s,b)$ respectively, $\ell_i$ are charged leptons $(e,\mu,\tau)$, $\nu_i$ are active left-handed (LH) neutrinos $(\nu_e,\nu_\mu,\nu_\tau)$, and $N_i$ are RH neutrinos. The quark fields and the charged lepton fields are in the mass basis, while the LH and RH neutrino fields are in the flavor basis.
Both $\nu_i$ and $N_i$ carry lepton number $L(\nu_i)=L(N_i)=+1$.
The flavors of the two quarks and those of the two leptons in the above operators can be different although we do not specify their flavor indexes here.
The dim-6 quark-lepton charged current operators which induce lepton number violation are
\begin{align}
\calO_{du\ell\nu1}^{V}=&(\overline{d_{L}}\gamma^\mu u_{L})(\overline{\ell_{R}^C}\gamma_\mu  \nu)\;, &
\calO_{du\ell\nu2}^{V}=&(\overline{d_{R}}\gamma^\mu u_{R})(\overline{\ell_{R}^C}\gamma_\mu  \nu)\;,
\nonumber\\
\calO_{du\ell\nu1}^{S}=&(\overline{d_{R}}u_{L})(\overline{\ell_{L}^C}\nu)\;,&
\calO_{du\ell\nu2}^{S}=&(\overline{d_{L}}u_{R})(\overline{\ell_{L}^C}\nu)\;,
\nonumber\\
\calO_{du\ell\nu}^{T}=&(\overline{d_{R}}\sigma^{\mu\nu}u_{L})(\overline{\ell_{L}^C}\sigma_{\mu\nu} \nu)\;,
\label{ope:dulNu}
\\
    \calO_{du\ell N1}^V=&(\overline{d_L}\gamma^\mu u_L)(\overline{\ell_L^C}\gamma_\mu N)\;,
&
\calO_{du\ell N2}^V=&(\overline{d_R}\gamma^\mu u_R)(\overline{\ell_L^C}\gamma_\mu N)\;,
\nonumber\\
\calO_{du\ell N1}^S=&(\overline{d_L}u_R)(\overline{\ell_R^C}N)\;,
&
\calO_{du\ell N2}^S=&(\overline{d_R}u_L)(\overline{\ell_R^C}N)\;,
\nonumber\\
\calO_{du\ell N}^T=&(\overline{d_L}\sigma^{\mu\nu} u_R)(\overline{\ell_R^C}\sigma_{\mu\nu} N)\;,
\label{ope:dulN}
\end{align}
together with their hermitian conjugates. In the above, charge conjugation of a fermion field $\psi$ is defined through $\psi^C=C\overline{\psi}^{\rm T}$ with the charge conjugation matrix $C$ satisfying $C^{\rm T}=C^\dagger=-C$ and $C^2=-1$. Next, we can incorporate the RG running effect from the experimental scale taken as the chiral symmetry breaking scale $\Lambda_{\chi}\simeq 1$ GeV to the electroweak scale $\Lambda_{\rm EW}\simeq m_W$ to match the LNEFT to the SMNEFT.

The RG equations for the Wilson coefficients of the charged currents in Eqs.~(\ref{ope:udlNu}-\ref{ope:dulN})  from 1-loop QCD and QED corrections are~\cite{Jenkins:2017dyc}
\begin{align}\nonumber
\mu{d\over d\mu}C^V_A=&+3{\alpha \over 2\pi}Q_uQ_eC^V_A\;, ~~~~~~~~~~~~~~~~~~~~~~~
C^V_A\in \left\{C_{ud\ell\nu1}^{V,prst},C_{ud\ell N2}^{V,prst},C_{du\ell\nu2}^{V,prst},C_{du\ell N1}^{V,prst}\right\}\;,
\\\nonumber
\mu{d\over d\mu}C^V_B=&-3{\alpha \over 2\pi}Q_dQ_eC^V_B\;, ~~~~~~~~~~~~~~~~~~~~~~~
C^V_B\in \left\{C_{ud\ell\nu2}^{V,prst},C_{ud\ell N1}^{V,prst},C_{du\ell\nu1}^{V,prst},C_{du\ell N2}^{V,prst}\right\}\;,
\\\nonumber
\mu{d\over d\mu}C^S_A=&-3\left({\alpha \over 2\pi}Q_uQ_d+{\alpha_s \over 2\pi}C_F \right)C^S_A\;,  ~~
C^S_A\in \left\{C_{ud\ell\nu2}^{S,prst},C_{ud\ell N2}^{S,prst},C_{du\ell\nu2}^{S,prst},C_{du\ell N2}^{S,prst}\right\}\;,
\\\nonumber
\mu{d\over d\mu}C^S=&-3\left({\alpha \over 2\pi}Q_uQ_d+{\alpha_s \over 2\pi}C_F \right)C^S+12{\alpha \over 2\pi}\left( Q_u^2 -Q_d^2\right)C^T\;,
\\
\mu{d\over d\mu}C^T=&+\left({\alpha \over 2\pi}(Q_u^2+Q_uQ_e-2Q_e^2)+{\alpha_s \over 2\pi}C_F \right)C^T+{1\over4}{\alpha \over 2\pi}\left( Q_u^2 -Q_d^2\right)C^S\;,
\label{RGeq}
\end{align}
with
\begin{align}
\begin{bmatrix}
C^S \\ C^T
\end{bmatrix}
\in
\left\{
\begin{bmatrix}
C_{ud\ell\nu1}^{S,prst} \\
C_{ud\ell\nu}^{T,prst}
\end{bmatrix},
\begin{bmatrix}
C_{ud\ell N1}^{S,prst} \\
C_{ud\ell N}^{T,prst}
\end{bmatrix},
\begin{bmatrix}
C_{du\ell\nu1}^{S,prst} \\
-C_{du\ell\nu}^{T,prst}
\end{bmatrix},
\begin{bmatrix}
C_{du\ell N1}^{S,prst} \\
-C_{du\ell N}^{T,prst}
\end{bmatrix}
\right\}\;,
\end{align}
where the electric charges $Q_u=2/3, Q_d=-1/3$ and $Q_e=-1$ for the up-type quarks, down-type quarks and charged leptons, respectively,
$C_F=(N_c^2-1)/2N_c$ with $N_c=3$ is the second Casimir operator of $SU(3)_c$ and
$\alpha={e^2\over 4\pi} (\alpha_s={g_s^2\over 4\pi})$ with $e (g_s)$ being the QED (QCD) coupling constant.
From the above RG equations we see the scalar-type operator ${O}_{qqe\nu(N)1}^S$ and tensor operator ${O}_{qqe\nu(N)}^T$ mix under the QED correction. The solution of the RG equations is given by
\begin{align}\nonumber
C^V_A(\mu_1)=&\left[\alpha_{\rm em}(\mu_2)\over \alpha_{\rm em}(\mu_1) \right]^{-3Q_uQ_e/ b_{e}}C^V_A(\mu_2)\;,
\\\nonumber
C^V_B(\mu_1)=&\left[\alpha_{\rm em}(\mu_2)\over \alpha_{\rm em}(\mu_1) \right]^{3Q_dQ_e/ b_{e}}C^V_B(\mu_2)\;,
\\
C^S_A(\mu_1)=&\left[\alpha_{s}(\mu_2)\over \alpha_{s}(\mu_1) \right]^{3C_F/ b}\left[\alpha_{\rm em}(\mu_2)\over \alpha_{\rm em}(\mu_1) \right]^{3Q_uQ_d/ b_{e}}C^S_A(\mu_2)\;,
\end{align}
in terms of the Wilson coefficients at the scales $\mu_1$ and $\mu_2$.
We introduced the coefficient $b=-11+2/3n_f$ from the RG equation of the QCD coupling with $n_f$ being the number of active quark flavors between scales $\mu_1$ and $\mu_2$, and the corresponding coefficient $b_e = \sum_i \tfrac43 (N_c)_i Q_i^2=4(3 n_\ell + 4 n_u +  n_d)/9$ from the RG equation of the QED coupling with $n_{\ell,u,d}$ being the active number of leptons/up-type quarks/down-type quarks between the two scales. For $C^S$ and $C^T$, the coupled differential equations have no analytical solutions.
We take the scale $\mu_1$ ($\mu_2$) to be  $\Lambda_\chi\simeq 1~\rm GeV$ ($\Lambda_{\text{EW}}\simeq m_W$), and use the 4-loop QCD running implemented in RunDec~\cite{Chetyrkin:2000yt} for $\alpha_s$ with initial value $\alpha_s(m_Z)=0.1179$
and the 1-loop QED running result for $\alpha$ with initial value $\alpha(m_e)=1/137.036$. After including quark and lepton threshold effects, the numerical results are
\begin{align}\nonumber
  C^V_A(\Lambda_\chi)=&1.01 C^V_A(\Lambda_{\rm EW})\;, ~~~~
  C^V_B(\Lambda_\chi)=1.01C^V_B(\Lambda_{\rm EW})\;, ~~~~
  C^S_A(\Lambda_\chi)=1.78C^S_A(\Lambda_{\rm EW})\;,
\\\nonumber
C^S(\Lambda_\chi)=&1.78 C^S(\Lambda_{\rm EW})-2.90\times 10^{-2} C^T(\Lambda_{\rm EW})\;,
\\
C^T(\Lambda_\chi)=&-5.16\times10^{-4}C^S(\Lambda_{\rm EW})+0.835 C^T(\Lambda_{\rm EW})\;.
\label{RG:ST}
\end{align}
We see the scalar Wilson coefficient $C^S(\Lambda_\chi)$ defined at $\Lambda_\chi$ receives a relatively considerable contribution from the tensor Wilson coefficient $C^T(\Lambda_{\rm EW})$ defined at $\Lambda_{\rm EW}$ via the mixing in RG running. For the low-energy observables sensitive to scalar operators but not the tensor operators, this mixing term is important to constrain the tensor Wilson coefficients through the direct constraints on scalar Wilson coefficients.

\subsection{Matching to the SMNEFT}
\label{sec:SMNEFT}
SMNEFT describes NP which enters at a sufficiently high scale above the electroweak scale. See Appendix~\ref{sec:SMNEFTbasis} for a complete list of SMNEFT operators involving RH neutrinos $N$ up to dim-7 and the relevant dim-6 and dim-7 operators without $N$. LNEFT should be matched to SMNEFT at the electroweak scale $\mu=m_W$ in order to constrain NP. We list the relevant tree-level matching conditions for the LNC and LNV cases in Table~\ref{matching}, and part of the matching results has also been given in Refs.~\cite{Jenkins:2017jig,Chala:2020vqp,Liao:2020zyx}.
Here $v=(\sqrt{2} G_F)^{-1/2}\simeq 246\, \mathrm{GeV}$ is the SM Higgs vacuum expectation value (vev), and $V$ is the unitary matrix transforming left-handed up-type quarks between flavor eigenstate $u_L^\prime$ and mass eigenstate $u_L$, i.e. $u_L^\prime=V^\dagger u_L$. We choose the basis, where the flavor and mass eigenstates of the charged leptons, the left-handed down-type quarks and RH quarks are identical. Thus the unitary matrix $V$ is the CKM matrix.

\begin{table}
\centering
\resizebox{\linewidth}{!}{
\renewcommand{\arraystretch}{1.2}
\begin{tabular}{|c| l l |}
\hline
Class & \multicolumn{2}{|c|}{Matching of the Wilson coefficients at the electroweak scale $\Lambda_{\rm EW}$}
\\
\hline
\hline
LNC& $C_{ud\ell\nu1}^{V,prst}=2V_{xr}C^{(3),tsxp*}_{lq}-{g_2^2 \over 2m_W^2}[W_l]_{ts}^*[W_q]_{pr} $
&
$C_{ud\ell\nu2}^{V,prst}=-{g_2^2 \over 2m_W^2}[W_l]_{ts}^*[W_R]_{pr}$
\\
$\ell\nu$ case & $C_{ud\ell\nu1}^{S,prst}=V_{xr}C_{lequ}^{(1),tsxp*} $
& $C_{ud\ell\nu2}^{S,prst}=C_{ledq}^{tsrp*}$
\\
& $C_{ud\ell\nu}^{T,prst}= V_{xr} C_{lequ}^{(3),tsxp*}$
&
\\
\hline
LNC& $C_{ud\ell N1}^{V,prst}=-{g_2^2 \over 2m_W^2}[W_q]_{pr} [W_{N}]_{ts}^*$
&
$C_{ud\ell N2}^{V,prst}=C_{duNe}^{rpts*}-{g_2^2 \over 2m_W^2}[W_R]_{pr} [W_{N}]_{ts}^* $
\\
$\ell N$ case & $C_{ud\ell N1}^{S,prst}=-C_{LNQd}^{stpr} +{1\over 2}C_{LdQN}^{srpt} $
&  $C_{ud\ell N2}^{S,prst}=V_{xr}C_{QuNL}^{xpts*}$
\\
 & $C_{ud\ell N}^{T,prst}= {1\over 8}C_{LdQN}^{srpt}$
 &
 \\
\hline
\hline
LNV& $C_{du\ell\nu1}^{V,prst}=+{g_2^2 \over 2m_W^2}[W_q]_{rp}^*[W_{\slashed l}]_{ts}$
&
$C_{du\ell\nu2}^{V,prst}=-{v \over  \sqrt{2} }C_{\bar duLeH}^{prts}+{g_2^2 \over 2m_W^2}[W_R]_{rp}^*[W_{\slashed l}]_{ts}$
\\
$\ell\nu$ case & $C_{du\ell\nu1}^{S,prst}=+{v \over  \sqrt{2}}C_{\bar dQLLH1}^{prst}$
&  $C_{du\ell\nu2}^{S,prst}=+{v \over  \sqrt{2}}V_{xp}^*C_{\bar QuLLH}^{xrst}$
\\
 & $C_{du\ell\nu}^{T,prst}=+{v \over  \sqrt{2}}C_{\bar dQLLH2}^{prst}$
&
\\
\hline
LNV& $C_{du\ell N1}^{V,prst}=-{v \over \sqrt{2}}V_{xp}^*C_{QNLH2}^{xrts}+{g_2^2 \over 2m_W^2}[W_q]_{pr}[W_{\slashed N}]_{ts} $
&
$C_{du\ell N2}^{V,prst}=+{v \over \sqrt{2}}C_{duNLH}^{prts}+{g_2^2 \over 2m_W^2}[W_R]_{pr}[W_{\slashed N}]_{ts}$
\\
$\ell N$ case & $C_{du\ell N1}^{S,prst}=+{v \over \sqrt{2}}V_{xp}^*C_{QuNeH1}^{xrts} $
&  $C_{du\ell N2}^{S,prst}=+{v \over \sqrt{2}}C_{dQNeH}^{prts} $
\\
 & $C_{du\ell N}^{T,prst}=-{v \over \sqrt{2}}V_{xp}^*C_{QuNeH2}^{xrts}$
 &
 \\
\hline
\end{tabular}
}
\caption{The matching result of the LNEFT and SMNEFT at the electroweak scale $\Lambda_{\rm EW}$.
The corresponding operators associated with the above SMNEFT Wilson coefficients are collected in Appendix~\ref{sec:SMNEFTbasis}.}
\label{matching}
\end{table}%

In Table~\ref{matching}, the modified coupling constants $[W_i]_{pr}$ of the $W$ boson with various charged currents are defined via the generalized charged current
\begin{align}\nonumber
{\cal L}\supset&
-{g_2 \over \sqrt{2}}W^{+\mu}\left(
[W_q]_{pr}\overline{u_{Lp}}\gamma_\mu d_{Lr}
+[W_R]_{pr}\overline{u_{Rp}}\gamma_\mu d_{Rr}
+[W_l]_{pr}\overline{\nu_p}\gamma_\mu \ell_{Lr}
+[W_N]_{pr}\overline{N_p}\gamma_\mu \ell_{Rr}\right)
\\
	       &-{g_2 \over \sqrt{2}}W^{+\mu}\left(
[W_{\slashed l}]_{pr}\overline{\nu_p^C}\gamma_\mu \ell_{Rr}
+[W_{\slashed N}]_{pr}\overline{N_p^C}\gamma_\mu \ell_{Lr}
\right)+{\rm h.c.}\;,
\end{align}
where the terms in the first (second) line conserve (violate) lepton number.
The dim-6 and dim-7 SMNEFT interactions shift the coupling constants to take the following form
\begin{align}\nonumber
[W_{q}]_{pr}=&V_{pr}+v^2C_{Hq}^{(3),px}V_{xr}\;, &
[W_{R}]_{pr}=&{1\over 2}v^2C_{Hud}^{pr}\;,
\\\nonumber
[W_{l}]_{pr}=&\delta_{pr}+v^2C_{Hl}^{(3),pr}\;, &
[W_{N}]_{pr}=&{1\over 2}v^2C_{HNe}^{pr}\;,
\\
[W_{\slashed l}]_{pr}=&-{v^3 \over 2\sqrt{2}}C_{LeHD}^{pr}\;,    &
[W_{\slashed N}]_{pr}=&-{v^3 \over 2\sqrt{2}}C_{NL1}^{pr}\;.
\end{align}
From the definition above and Table~\ref{matching}, we see the SM contribution enters the matching only through the LNC operator ${\cal O}_{ud\ell\nu1}^V$ with
Wilson coefficient $C_{ud\ell\nu 1,\rm SM}^{V, pr\alpha\beta}=-{4G_F\over \sqrt{2}}V_{pr}\delta_{\alpha\beta}$. For the operators ${\cal O}_{ud\ell N2}$, ${\cal O}_{du\nu e2}$, ${\cal O}_{du\ell N2}$, the matching contribution from integrating out the $W$ boson is bilinear in terms of the SMNEFT dim-6 and dim-7 Wilson coefficients, therefore, the contribution is doubly suppressed and can be neglected in the numerical analysis.

\section{Low-energy processes and relevant constraints}
\label{sec:Const}

In this section we discuss the sensitivity of low-energy precision measurements to the charged currents in LNEFT and give the analytical expressions for the constraints. We consider decay branching fractions, the LFU and the CKM unitarity in leptonic decays of pseudoscalar mesons and hadronic tau lepton decays. Below, we first present the general expressions of the partial decay widths needed for the observables.

We restrict ourselves to 2-body decays and calculate the decay widths $\Gamma(M^+\to \ell^+ \nu)$ and $\Gamma(\tau^-\to M^- \nu)$ with $M$ denoting a light meson. We denote the meson $M$ as $M^+(u_p\bar d_r)\equiv M^+$ and $M^-(\bar u_p d_r)\equiv M^-$ and it can be either a pseudoscalar meson $P$ or a vector meson $V$.
For a pseudoscalar meson $P$, the transition matrix elements to the vacuum state from the scalar, vector, and tensor quark currents are zero. The only non-vanishing matrix elements are for pseudoscalar currents, axial-vector currents and the anomaly matrix elements. The relevant decay constants for our study are~\cite{Beneke:2002jn,Cheng:2013fba}
\begin{align}
  \left\langle0|\bar q\gamma^\mu\gamma_5q^\prime|P(p)\right\rangle&=i f_P p^\mu\;,
&
  \left\langle0|\bar q\gamma_5q^\prime|P(p)\right\rangle&=-i {h_P\over {m_q+m_{q^\prime}}} \;,
\end{align}
where $f_P$ is the decay constant of pseudoscalar $P$ and, for different-flavor quarks $q\neq q^\prime$, $h_P=m_P^2 f_P$ with $m_P$ being the mass of $P$.
The non-vanishing hadronic matrix elements for a vector meson $V$ with momentum $p$ and polarization vector $\epsilon^\mu_V$ can be parameterized as~\cite{Ball:2006eu,Cheng:2013fba}
\begin{align}
\langle 0|\bar{q}\gamma^\mu q^\prime|V(p)\rangle =& f_V m_V \epsilon^\mu_V \; , &
\langle 0|\bar{q}\sigma^{\mu\nu} q^\prime|V(p)\rangle =& if_V^{T} \left(\epsilon^\mu_V p^\nu - \epsilon^\nu_Vp^\mu \right) \; .
\end{align}
where $f_V$ and $f_V^T$ are the vector and tensor form factors pertinent to the vector meson $V$, and $m_V$ is the mass of $V$.
In the following we assume the limit of massless right-handed neutrinos.

For the charged leptonic decay of a pseudoscalar meson $P^+$, from Eqs.~(\ref{ope:udlNu}, \ref{ope:udlN}), we obtain the following $\Delta L=0$ partial decay widths
\begin{eqnarray}
\Gamma(P^+\to \ell^+_\alpha X_\beta)={(m_P^2-m_{\ell_\alpha}^2)^2m_{\ell_\alpha}^2f_P^2\over 64\pi m_P^3}\Big|C_{ud\ell X 1}^{V}-C_{ud\ell X 2}^{V} - \eta_X J_{P\ell_\alpha}(C_{ud\ell X 1}^{S}-C_{ud\ell X 2}^{S})\Big|^2\;,
\end{eqnarray}
where $X\in \{\nu,N\}$ and we defined $\eta_\nu=1$ and $\eta_N=-1$.
We do not show the flavor indices of the Wilson coefficients explicitly for brevity. They are easily recognized in terms of the quark content of the pseudoscalar meson $P^+$ and the flavors of charged lepton $\ell$ and neutrino $\nu/N$ (the same abbreviated convention is also implied in the following unless we show them explicitly for specific observables). $m_{\ell_\alpha}$ is the mass of the charged lepton $\ell_\alpha$ and $J_{P\ell_\alpha}$ denotes the mass ratio
\begin{align}
J_{P\ell_\alpha}={m_P^2 \over m_{\ell_\alpha}(m_{u_p}+m_{d_r})}\;.
\end{align}
The Wilson coefficients $C_{ud\ell\nu1}^V$ with the same-lepton-flavor components can be split into the SM and NP contributions, i.e. $C_{ud\ell\nu 1}^{V, pr\alpha\alpha}=C_{ud\ell\nu 1,\rm  SM}^{V, pr\alpha\alpha}+C_{ud\ell\nu 1,\rm NP}^{V, pr\alpha\alpha}$ with $C_{ud\ell\nu 1,\rm SM}^{V, pr\alpha\beta}=-{4G_F\over \sqrt{2}}V_{pr}\delta_{\alpha\beta}$.
In the SM, only the flavor conserving transition $P^+\to \ell_\alpha^+\nu_\alpha$ is allowed and the decay width takes the form
\begin{eqnarray}
\Gamma(P^+\to \ell^+_\alpha \nu_\alpha)_{\rm SM}={G_F^2|V_{pr}|^2 f_P^2 m_{\ell_\alpha}^2 m_P\over 8\pi}\left(1-\frac{m_{\ell_\alpha}^2}{m_P^2}\right)^2(1+\delta_P)\;,
\label{SM:plnu}
\end{eqnarray}
where the flavor indices $p,r$ in the CKM matrix correspond to the two quarks constituting the meson and $\delta_P$ parameterizes radiative corrections. From Eqs.~(\ref{ope:dulNu}, \ref{ope:dulN}), the
$\Delta L=-2$ partial decay widths with a final state anti-neutrino $\bar X\in\{\bar\nu,\bar N\}$ are
\begin{eqnarray}
\Gamma(P^+\to \ell^+_\alpha \bar{X}_\beta)&=&{(m_P^2-m_{\ell_\alpha}^2)^2m_{\ell_\alpha}^2f_P^2\over 64\pi m_P^3}\Big|C_{du\ell X 1}^{V}-C_{du\ell X 2}^{V}+\eta_X J_{P\ell_\alpha} (C_{du\ell X 1}^{S}-C_{du\ell X 2}^{S})\Big|^2\;.
\end{eqnarray}
For tau lepton decay into a pseudoscalar meson and a neutrino, we have the following $\Delta L=0$ and $\Delta L=-2$ partial decay widths
 \begin{eqnarray}
\Gamma(\tau^-\to P^- X_\beta)&=&{(m_\tau^2-m_P^2)^2f_P^2 \over 128\pi m_\tau}\Big|C_{ud\ell X 1}^{V}-C_{ud\ell X 2}^{V}-\eta_X J_{P\tau}(C_{ud\ell X 1}^{S}-C_{ud\ell X 2}^{S})\Big|^2\;,\\
\Gamma(\tau^-\to P^-\bar{X}_\beta)&=&{(m_\tau^2-m_P^2)^2f_P^2\over 128\pi m_\tau}\Big|C_{du\ell X 1}^{V}-C_{du\ell X 2}^{V}+\eta_X J_{P\tau} (C_{du\ell X 1}^{S}-C_{du\ell X 2}^{S})\Big|^2\;,
\end{eqnarray}
and the SM partial width is
\begin{eqnarray}
  \Gamma(\tau^-\to P^- \nu_\tau)_{\rm SM}= {G_F^2|V_{pr}|^2 f_P^2 m_\tau^3\over 16\pi}\left(1-\frac{m_P^2}{m_\tau^2}\right)^2 (1+\delta_{\tau P})\;,
\label{SM:lpnu}
\end{eqnarray}
where $\delta_{\tau P}$ describes radiative corrections.
The partial decay widths of tau lepton decay into a vector meson and a neutrino are
\begin{eqnarray}
\Gamma(\tau^-\to V^- X_\beta)&=& {(m_\tau^2-m_V^2)^2\over 32\pi m_\tau^3} \Big[{1\over 4}|C^V_{ud\ell X 1}+C^V_{ud\ell X 2}|^2 (f_V)^2 (m_\tau^2+2m_V^2)  \\\nonumber
			     &+&4|C_{ud\ell X}^T|^2 (f_V^{T})^2 (2m_\tau^2+m_V^2)
+ 6 {\rm Re}[(C^V_{ud\ell X 1}+C^V_{ud\ell X 2})C^{T\ast}_{ud\ell X}]m_\tau m_V f_V f_V^{T}
\Big] \;, \\
\Gamma(\tau^-\to V^- \bar{X}_\beta)&=& {(m_\tau^2-m_V^2)^2\over 32\pi m_\tau^3} \Big[{1\over 4}|C^V_{du\ell X 1}+C^V_{du\ell X 2}|^2 (f_V)^2 (m_\tau^2+2m_V^2) \\\nonumber
				   &+&4|C_{du\ell X}^T|^2 (f_V^{T})^2 (2m_\tau^2+m_V^2)
- 6 {\rm Re}[(C^V_{du\ell X 1}+C^V_{du\ell X 2})C^{T\ast}_{du\ell X}]m_\tau m_V f_V f_V^{T}
\Big] \;.
\end{eqnarray}
The partial width in the SM is
\begin{eqnarray}
\Gamma(\tau^-\to V^- \nu_\tau)_{\rm SM}&=& {G_F^2 m_\tau |V_{pr}|^2\over 8\pi}(f_V)^2 m_V^2 \Big(1+{m_\tau^2\over 2m_V^2} \Big) \Big(1-{m_V^2\over m_\tau^2} \Big)^2\;.
\end{eqnarray}
The matrix elements of all above processes are collected in Appendix~\ref{sec:mesonmatrix}.

To evaluate the constraints on the LNEFT WCs from low-energy measurements, we need to define a quantity for each observable in experiment and suppose a general observable quantity $\mathcal{Q}$ below. We are able to assume that the experimental error $\delta \mathcal{Q}_{\rm exp}$ and theoretical error in the SM calculation $\delta \mathcal{Q}_{\rm SM}$ are uncorrelated and thus simply use Gaussian error propagation to obtain the uncertainty with respect to the SM prediction
\begin{align}
\delta\left(\frac{\mathcal{Q}_{\rm exp}}{\mathcal{Q}_{\rm SM}}\right)
= \left[
\left(\frac{\delta \mathcal{Q}_{\rm exp}}{\mathcal{Q}_{\rm exp}^{\rm cent.}}\right)^2 +
\left(\frac{\delta \mathcal{Q}_{\rm SM}}{\mathcal{Q}_{\rm SM}^{\rm cent.}}\right)^2 \right]^{1\over 2} \left|\frac{\mathcal{Q}_{\rm exp}^{\rm cent.}}{\mathcal{Q}_{\rm SM}^{\rm cent.}}\right|\;,
\label{error}
\end{align}
where $\mathcal{Q}^{\rm cent.}_{\rm exp}$ and $\mathcal{Q}^{\rm cent.}_{\rm SM}$ denote the central values of experimental measurement and theoretical prediction, respectively.
For the observable ${\cal Q}$, we denote the
NP contribution encoded in the effective operators relative to SM prediction by $\Delta R_{\cal Q}={ {\cal Q}_{\rm EFT}\over {\cal Q}_{\rm SM}}-1$ and demand that ${ {\cal Q}_{\rm EFT}\over {\cal Q}_{\rm SM}}$ is compatible with the experimental result over the SM value ${ {\cal Q}_{\rm exp}\over {\cal Q}_{\rm SM}}={ {\cal Q}_{\rm exp}^{\rm cent.}\over {\cal Q}_{\rm SM}^{\rm cent.}}\pm \delta\left( { {\cal Q}_{\rm exp}\over {\cal Q}_{\rm SM}}\right)$. Then we have the constraint on the deviation $\Delta R_{\cal Q}$
\begin{align}
\left| \Delta R_{\cal Q}\right|\leq \left|1-{ {\cal Q}_{\rm exp}^{\rm cent.}\over {\cal Q}_{\rm SM}^{\rm cent.}}\right|+\delta\left( { {\cal Q}_{\rm exp}\over {\cal Q}_{\rm SM}}\right)\;.
\label{eq:bound}
\end{align}
Next we explore each observable in terms of LNEFT WCs and evaluate the corresponding errors.

\subsection{Decay branching ratios}
\label{sec:BR}

\begin{table}
\centering
\begin{tabular}{|c|c||c|c|}
\hline
Decay & BR & Decay & BR
\\\hline
$\pi^+\to \mu^+ \nu_\mu$ & $(99.98770\pm 0.00004)\%$ & $K^+\to \mu^+ \nu_\mu$ & $(63.56\pm 0.11)\%$
\\
$\pi^+\to e^+ \nu_e$ & $(1.230\pm 0.004)\times 10^{-4}$ & $K^+\to e^+ \nu_e$ & $(1.582\pm 0.007)\times 10^{-5}$ at 90\% CL
\\
$\pi^+\to \mu^+ \bar{\nu}_e$ & $<1.5\times 10^{-3}$  at 90\% CL & $K^+\to \mu^+ \bar{\nu}_e$ & $<3.3\times 10^{-3}$ at 90\% CL
\\
$\pi^+\to \mu^+ \nu_e$ & $<8.0\times 10^{-3}$ at 90\% CL & $K^+\to \mu^+ \nu_e$ & $<4.0\times 10^{-3}$ at 90\% CL
\\\hline
$\tau^-\to \pi^- \nu_\tau$ & $(10.82\pm 0.05)\%$ & $\tau^-\to K^- \nu_\tau$ & $(6.96\pm 0.10)\times 10^{-3}$
\\
$\tau^-\to \rho^- \nu_\tau$ & $(25.2\pm0.4)\times 10^{-2}$ & $\tau^-\to K^{*-} \nu_\tau$ & $(1.20\pm0.07)\times 10^{-2}$
\\\hline
\end{tabular}
\caption{The experimental decay branching fractions of the relevant leptonic decays of pseudoscalar mesons and hadronic tau lepton decays~\cite{Tanabashi:2018oca}.
}
\label{tab:BR}
\end{table}

The first observable is the decay branching fraction of pseudoscalar mesons and hadronic tau lepton decays. The observable quantity is the partial decay width and the experimental results of decay branching fractions are given in Table~\ref{tab:BR}.
For the charged leptonic decay of a pseudoscalar meson $P^+$, we obtain the following partial decay width with respect to the SM prediction
\begin{eqnarray}
  1+\Delta_{P\ell_\alpha} & \equiv &{\Gamma(P^+\to \ell^+_\alpha +{\rm inv.})\over \Gamma(P^+\to \ell^+_\alpha\nu_\alpha)_{\rm SM}}\;,
 \label{EFT:plinv}
\\\nonumber
\Delta_{P\ell_\alpha}&\simeq&
2 {\rm Re}\left[(C^{V,pr\alpha\alpha}_{ud\ell\nu 1,\rm  SM})^{-1} \left(C^{V,pr\alpha\alpha}_{ud\ell\nu 1,\rm  NP}-C^{V,pr\alpha\alpha}_{ud\ell\nu 2}- J_{P\ell_\alpha}(C^{S,pr\alpha\alpha}_{ud\ell\nu 1}-C^{S,pr\alpha\alpha}_{ud\ell\nu 2})\right)\right]
\\\nonumber
&+&{1\over |C^{V,pr\alpha\alpha}_{ud\ell\nu 1,\rm SM}|^2}\sum_{\beta\neq \alpha}\Big|C_{ud\ell\nu 1}^{V,pr\alpha\beta}-C_{ud\ell\nu 2}^{V,pr\alpha\beta}- J_{P\ell_\alpha}(C_{ud\ell\nu 1}^{S,pr\alpha\beta}-C_{ud\ell\nu 2}^{S,pr\alpha\beta})\Big|^2
\\\nonumber
&+&{1\over |C^{V,pr\alpha\alpha}_{ud\ell\nu 1,\rm SM}|^2}\sum_{\beta}\Big|C_{ud\ell N 1}^{V,pr\alpha\beta}-C_{ud\ell N 2}^{V,pr\alpha\beta}+J_{P\ell_\alpha}(C_{ud\ell N 1}^{S,pr\alpha\beta}-C_{ud\ell N 2}^{S,pr\alpha\beta})\Big|^2
\\\nonumber
&+&{1\over |C^{V,pr\alpha\alpha}_{ud\ell\nu 1,\rm SM}|^2}\sum_{\beta}\Big|C_{du\ell\nu 1}^{V,rp\alpha\beta}-C_{du\ell\nu 2}^{V,rp\alpha\beta}+J_{P\ell_\alpha} (C_{du\ell\nu 1}^{S,rp\alpha\beta}-C_{du\ell\nu 2}^{S,rp\alpha\beta})\Big|^2
\\
&+&{1\over |C^{V,pr\alpha\alpha}_{ud\ell\nu 1,\rm SM}|^2}\sum_{\beta}\Big|C_{du\ell N 1}^{V,rp\alpha\beta}-C_{du\ell N 2}^{V,rp\alpha\beta}-J_{P\ell_\alpha}(C_{du\ell N 1}^{S,rp\alpha\beta}-C_{du\ell N 2}^{S,rp\alpha\beta})\Big|^2\;,
\end{eqnarray}
where $\beta$ labels the flavor of either active neutrino $\nu$ or sterile neutrino $N$.

For pseudoscalar meson $P=\pi, K$ decay, the theoretical uncertainties are from both the radiative
correction factor  $\delta_{P}$ and the decay constant $f_P$. They are $\delta_\pi=0.0176\pm 0.0021$ and $f_\pi=130.2\pm 1.2~{\rm MeV}$ for the pion and $\delta_K=0.0107\pm 0.0021$ and $f_K=155.7\pm 0.3~{\rm MeV}$ for the Kaon~\cite{Tanabashi:2018oca,Aoki:2019cca}.
The experimental uncertainties are from both the lifetime measurement and the BRs in Table~\ref{tab:BR}. The lifetime of the charged pion and Kaon are $\tau_{\pi^\pm}=26.033\pm0.005$ ns and $\tau_{K^\pm}=12.38\pm 0.02$ ns, respectively~\cite{Tanabashi:2018oca}.
For the lepton-number/flavor-conserving (LNC/LFC) decays of $\pi^+$ and $K^+$, the theoretical and experimental uncertainties together with Eq.~(\ref{error}), result in the following central values and errors for $\pi$ and $K$ decays
\begin{align}
{\Gamma_{\rm exp}^{\pi e}\over \Gamma_{\rm SM}^{\pi e}}=& 0.963\pm 0.018 \;,
&{\Gamma_{\rm exp}^{\pi \mu}\over \Gamma_{\rm SM}^{\pi \mu}}=& 1.005\pm 0.019\;,
\nonumber
\\
{\Gamma_{\rm exp}^{K e}\over \Gamma_{\rm SM}^{K e}}=& 0.9770\pm0.0063\;,
&{\Gamma_{\rm exp}^{K \mu}\over \Gamma_{\rm SM}^{K \mu}}=& 1.0084\pm0.0050\;.
\end{align}

In Table~\ref{tab:BR} we also show the limits on the branching fractions of $\pi$ and $K$ decay into $\mu^+$ and electron or anti-electron neutrino at
the 90\% confidence level (CL).
The relevant data were taken from the bubble chamber BEBC during the short baseline run of the CERN SPS $\nu$ and $\bar{\nu}$ wide band beams. No $\bar{\nu}_e$ excess has been found in $\nu_\mu \to \bar{\nu}_e$ and $\nu_e\to \bar{\nu}_e$ transitions and thus bounds have been placed on these LFV and LNV decays~\cite{CooperSarkar:1981pb}.
We use these bounds to constrain the $\mu,e$ component in the WCs contributing to the relevant processes.
For LFV decays $P^+\to \mu^+ \nu_e$ and LNV decays $P^+\to \mu^+ \bar{\nu}_e$, the partial decay widths are
\begin{eqnarray}
\Gamma(P^+\to \mu^+ \nu_{e})&=&{(m_P^2-m_{\mu}^2)^2 f_P^2 m_{\mu}^2\over 64\pi m_P^3}
\Big|C_{ud\ell\nu 1}^{V,pr\mu e}-C_{ud\ell\nu 2}^{V,pr\mu e}-J_{P\mu}(C_{ud\ell\nu 1}^{S,pr\mu e}-C_{ud\ell\nu 2}^{S,pr\mu e})\Big|^2\;,\\
\Gamma(P^+\to \mu^+ \bar{\nu}_{e})&=&{(m_P^2-m_{\mu}^2)^2 f_P^2 m_{\mu}^2\over 64\pi m_P^3}
\Big|C_{du\ell\nu 1}^{V,rp\mu e}-C_{du\ell\nu 2}^{V,rp\mu e}+J_{P\mu}(C_{du\ell\nu 1}^{S,rp\mu e}-C_{du\ell\nu 2}^{S,rp\mu e})\Big|^2\;.
\end{eqnarray}

For tau lepton decay into a pseudoscalar meson $P^-$ and a neutrino, the deviation due to NP is described by $\Delta_{\tau P}$
\begin{eqnarray}
  1+\Delta_{\tau P} &\equiv & {\Gamma(\tau^-\to P^- +{\rm inv.})\over \Gamma(\tau^-\to P^- \nu_\tau)_{\rm SM}}
\label{EFT:lpinv}
\\
  \Delta_{\tau P}
  &\simeq& 2 {\rm Re} \left[(C^{V,pr\tau\tau}_{ud\ell\nu 1,\rm SM})^{-1}\Big(C^{V,pr\tau\tau}_{ud\ell\nu 1,\rm NP}-C^{V,pr\tau\tau}_{ud\ell\nu 2}-J_{P\tau}(C^{S,pr\tau\tau}_{ud\ell\nu 1}-C^{S,pr\tau\tau}_{ud\ell\nu 2})\Big)\right] \nonumber \\
  &+&{1\over |C^{V,pr\tau\tau}_{ud\ell\nu 1,\rm SM}|^2}\sum_{\beta\neq\tau}\Big|C^{V,pr\tau\beta}_{ud\ell\nu 1}-C^{V,pr\tau\beta}_{ud\ell\nu 2}-J_{P\tau}(C^{S,pr\tau\beta}_{ud\ell\nu 1}-C^{S,pr\tau\beta}_{ud\ell\nu 2})\Big|^2\nonumber \\
  &+&{1\over |C^{V,pr\tau\tau}_{ud\ell\nu 1,\rm SM}|^2}\sum_{\beta}\Big|C^{V,pr\tau\beta}_{ud\ell N 1}-C^{V,pr\tau\beta}_{ud\ell N 2}+J_{P\tau}(C^{S,pr\tau\beta}_{ud\ell N 1}-C^{S,pr\tau\beta}_{ud\ell N 2})\Big|^2\nonumber \\
  &+&{1\over |C^{V,pr\tau\tau}_{ud\ell\nu 1,\rm SM}|^2}\sum_{\beta}\Big|C^{V,rp\tau\beta}_{du\ell\nu 1}-C^{V,rp\tau\beta}_{du\ell\nu 2}+J_{P\tau}(C^{S,rp\tau\beta}_{du\ell\nu 1}-C^{S,rp\tau\beta}_{du\ell\nu 2})\Big|^2\nonumber \\
  &+&{1\over |C^{V,pr\tau\tau}_{ud\ell\nu 1,\rm SM}|^2}\sum_{\beta}\Big|C^{V,rp\tau\beta}_{du\ell N 1}-C^{V,rp\tau\beta}_{du\ell N 2}-J_{P\tau}(C^{S,rp\tau\beta}_{du\ell N 1}-C^{S,rp\tau\beta}_{du\ell N 2})\Big|^2\;.
\end{eqnarray}
In the SM calculation of tau lepton's hadronic decay into a pseudoscalar meson and a neutrino, the main theoretical uncertainties are from meson decay constants $f_P$ and higher order corrections, i.e.
$\delta_{\tau\pi}=0.0192\pm 0.0024$ and
$\delta_{\tau K}=0.0198\pm 0.0031$~\cite{Decker:1994ea,Cirigliano:2007ga,Rosner:2015wva}.
The experimental uncertainties are from the tau lepton lifetime $\tau_{\tau^\pm}=(290.3\pm 0.5)\times10^{-6}~{\rm ns}$ and the decay branching fractions in Table~\ref{tab:BR}.
For $P^+=\pi^+, K^+$, the theoretical and experimental uncertainties result in the following central values and errors
\begin{align}
&{\Gamma_{\rm exp}^{\tau\pi}\over \Gamma_{\rm SM}^{\tau\pi}}= 0.998\pm  0.019\;,
&&{\Gamma_{\rm exp}^{\tau K}\over \Gamma_{\rm SM}^{\tau K}}= 0.980\pm 0.015\;.
\end{align}
For tau lepton's hadronic decay into a vector meson $V^-$, the deviation is given by
\begin{eqnarray}
1+\Delta_{\tau V} &\equiv & {\Gamma(\tau^-\to V^- +{\rm inv.})\over \Gamma(\tau^-\to V^- \nu_\tau)_{\rm SM}}\;,
\\
\Delta_{\tau V}
&\simeq& 2 {\rm Re} \left[(C^{V,pr\tau\tau}_{ud\ell\nu 1,\rm SM})^{-1}\Big(C^{V,pr\tau\tau}_{ud\ell\nu 1,\rm NP}+C^{V,pr\tau\tau}_{ud\ell\nu 2}+C^{T,pr\tau\tau}_{ud\ell\nu}K'_{\tau V} \Big)\right]
 \nonumber \\
  &+&{1\over |C^{V,pr\tau\tau}_{ud\ell\nu 1,\rm SM}|^2}\sum_{\beta}\Big( \Big|C^{V,pr\tau\beta}_{ud\ell\nu 1}+C^{V,pr\tau\beta}_{ud\ell\nu 2}\Big|^2_{\beta\neq \tau}
+\Big|C^{V,pr\tau\beta}_{ud\ell N 1}+C^{V,pr\tau\beta}_{ud\ell N 2}\Big|^2\Big)
\nonumber \\
  &+&{1\over |C^{V,pr\tau\tau}_{ud\ell\nu 1,\rm SM}|^2}\sum_{\beta}\Big(\Big|C^{V,rp\tau\beta}_{du\ell\nu 1}+C^{V,rp\tau\beta}_{du\ell\nu 2}\Big|^2+\Big|C^{V,rp\tau\beta}_{du\ell N 1}+C^{V,rp\tau\beta}_{du\ell N 2}\Big|^2\Big)
 \nonumber \\
&+&{K_{\tau V}\over |C^{V,pr\tau\tau}_{ud\ell\nu 1,\rm SM}|^2}\sum_{\beta}\Big(
\Big|C_{ud\ell\nu}^{T,pr\tau\beta}\Big|^2  + \Big|C_{ud\ell N}^{T,pr\tau\beta}\Big|^2 +\Big|C_{du\ell\nu}^{T,rp\tau\beta}\Big|^2+\Big|C_{du\ell N}^{T,rp\tau\beta}\Big|^2\Big)\;,
\end{eqnarray}
where we omit the interference terms among NP contributions but only keep the interference terms containing SM part shown in the first line above. The two dimensionless factors are defined as
\begin{align}
K_{\tau V}=&{16 (f_V^{T})^2 (2m_\tau^2+m_V^2)\over f_V^2 (m_\tau^2+2m_V^2)}\;,
&K'_{\tau V}=&{12 f_V^{T} m_\tau m_V\over f_V(m_\tau^2+2m_V^2)}\;.
\end{align}
For $V^-=\rho^-, K^{\ast -}$, the main theoretical uncertainties are from scalar and tensor decay constants $f_\rho^{(T)}$ and $f_{K^\ast}^{(T)}$~\cite{Becirevic:2003pn,Ball:2006eu}. In our analysis we use the numerical values for the form factors given in Ref.~\cite{Ball:2006eu}:
$f_\rho = (206\pm 7)~{\rm MeV}$,
$f_\rho^{T}(\mu=1~\mathrm{GeV})=(165\pm 9)~{\rm MeV}$,
$f_{K^\ast} = 222\pm 8~{\rm MeV}$, and
$f_{K^\ast}^{T}(\mu=1~\mathrm{GeV})=185\pm 10~{\rm MeV}$.
The experimental uncertainties are from tau lifetime and the measured branching ratios~\cite{Tanabashi:2018oca}.
We then obtain
\begin{align}
&{\Gamma_{\rm exp}^{\tau\rho}\over \Gamma_{\rm SM}^{\tau\rho}}= 1.032\pm 0.072\;, &&{\Gamma_{\rm exp}^{\tau K^\ast}\over \Gamma_{\rm SM}^{\tau K^\ast}}= 0.860\pm 0.080\;.
\end{align}

\subsection{LFU in pseudoscalar meson and $\tau$ lepton decays}
\label{sec:LFU}

\begin{table}
  \begin{tabular}{|c|c|c|}\hline
    Observable & exp. & SM theory\\\hline
    $R_{e/\mu}^{\pi}$ & $(1.2344\pm0.0030) \times 10^{-4}$~\cite{Aguilar-Arevalo:2015cdf,Mischke:2018qmv} & $(1.2352\pm0.0001)\times 10^{-4}$~\cite{Cirigliano:2007xi}\\
    $R_{e/\mu}^{K}$ & $(2.488\pm0.010) \times 10^{-5}$~\cite{Lazzeroni:2012cx}
 & $(2.477\pm0.001)\times 10^{-5}$~\cite{Cirigliano:2007xi}\\
    $(g_{\tau/\mu})_\pi$ & $0.9958\pm0.0026$~\cite{Amhis:2019ckw} & $1$\\
    $(g_{\tau/\mu})_K$ & $0.9879\pm0.0063$~\cite{Amhis:2019ckw} & $1$ \\\hline
  \end{tabular}
  \caption{LFU observables. Experimental measurements are summarized in the second column and the SM prediction in the third column.}
  \label{tab:LFU}
\end{table}

For LFU in pseudoscalar meson decay, we define the observable as the ratio of the decay widths~\cite{Pich:2013lsa}
\begin{eqnarray}
  R^P_{e/\mu}\equiv{\Gamma(P\to e+{\rm inv.})\over \Gamma(P\to \mu+{\rm inv.})} \simeq R^P_{e/\mu,\rm SM} \left[1+\Delta_{Pe}-\Delta_{P\mu}\right]
\;.
\end{eqnarray}
where we have factorized out the NP corrections encoded in $\Delta_{P\ell}$ relative to the SM prediction in the latter step.
The experimental LFU measurement and the SM predictions of the above ratio for $P=\pi, K$~\cite{Cirigliano:2007xi} are collected in Table~\ref{tab:LFU}.
We then obtain the combined error
\begin{align}
&{R^\pi_{e/\mu,\rm exp}\over R^\pi_{e/\mu,\rm SM}}=0.9994\pm  0.0024\;,
&& {R^K_{e/\mu,\rm exp}\over R^K_{e/\mu,\rm SM}}=1.0044\pm  0.0041\;.
\end{align}
For LFU in tau lepton decay, we define the observable as~\cite{Amhis:2019ckw}
  \begin{eqnarray}
    (g_{\tau/\mu})^2_{P}\equiv{\Gamma(\tau\to P+{\rm inv.}) 2m_P m_\mu^2\over \Gamma(P\to \mu+{\rm inv.})(1+\delta R_{\tau/P}) m_\tau^3} \Big({1-m_\mu^2/m_P^2\over 1-m_P^2/m_\tau^2}\Big)^2
\end{eqnarray}
with the radiative corrections being
$\delta R_{\tau/\pi}=(0.16\pm 0.14)\%$ and
$\delta R_{\tau/K}=(0.90\pm 0.22)\%$~\cite{Marciano:1993sh,Decker:1994dd,Decker:1994ea}.
The factor $1+\delta R_{\tau/P}$ compensates the radiative corrections $\delta_P$ and $\delta_{\tau P}$ in the SM partial widths and then $(g_{\tau/\mu})_{P, {\rm SM}}=1$.
Thus, the uncertainty of $(g_{\tau/\mu})_{P}$ only comes from the experimental uncertainty, which are quoted in Table~\ref{tab:LFU}.
We can then obtain for the ratio of the measured value with respect to the SM prediction
\begin{align}
&{(g_{\tau/\mu})_{\pi,{\rm exp}}\over (g_{\tau/\mu})_{\pi,{\rm SM}}}
=0.9958\pm 0.0026\;, \quad
&&{(g_{\tau/\mu})_{K,{\rm exp}}\over (g_{\tau/\mu})_{K,{\rm SM}}}
=0.9879\pm 0.0063 \;.
\end{align}
In the presence of NP contributions we find
  \begin{eqnarray}
(g_{\tau/\mu})_{P}\simeq
1+{1\over 2}\Delta_{\tau P}-{1\over 2}\Delta_{P\mu}
\end{eqnarray}
and it thus depends on both $\Delta_{\tau P}$ and $\Delta_{P\mu}$.

\subsection{CKM unitarity}
\label{sec:CKM}

\begin{table}
\centering
\begin{tabular}{|c|c|}
\hline
Decay & CKM
\\\hline
$\tau\to K\nu$ & $|V_{us}^{\tau K}|=0.2234\pm 0.0015$~\cite{Amhis:2019ckw}
\\
${\Gamma(\tau\to K \nu)\over \Gamma(\tau\to \pi \nu)}$ & $|V_{us}^{\tau K/\pi}|=0.2236\pm 0.0015$~\cite{Amhis:2019ckw}
\\
${\Gamma(K\to \mu \nu)\over \Gamma(\pi\to \mu \nu)}$ & $|V_{us}^{K/\pi\mu}|=
0.22535\pm0.00044$~\cite{Coutinho:2019aiy}
\\
nuclear $\beta$ (CMS) & $|V_{ud}^\beta|_{\rm CMS}=0.97389\pm 0.00018$~\cite{Czarnecki:2019mwq} \\
		       nuclear $\beta$ (SGPR) & $|V_{ud}^\beta|_{\rm SGR}=0.97370\pm 0.00014$~\cite{Seng:2018qru}
\\\hline
\end{tabular}
\caption{The CKM observables.
}
\label{tab:CKM}
\end{table}

Furthermore, NP contributions in the neutrino sector~\cite{Lusiani:2018zvr,Grossman:2019bzp,Coutinho:2019aiy,Czarnecki:2019iwz} may also affect the determination of the CKM matrix elements. Concretely, the NP contribution leads to a shift in the extracted CKM matrix elements. From Eqs.~(\ref{SM:plnu},~\ref{EFT:plinv}) and Eqs.~(\ref{SM:lpnu},~\ref{EFT:lpinv}), we find
\begin{align}
|V^{P\ell}_{pr}|\simeq&|V^{\cal L}_{pr}| \Big(1+{1\over 2}\Delta_{P\ell}\Big)\;, &
|V^{\tau P}_{pr}|\simeq&|V^{\cal L}_{pr}| \Big(1+{1\over 2}\Delta_{\tau P}\Big)\;,
\end{align}
where $V^{\cal L}_{pr}$ is the CKM matrix element in the SM Lagrangian, which satisfies the unitarity condition. In particular, the decays $\tau\to K\nu$, $\Gamma(\tau\to K\nu)/\Gamma(\tau\to \pi\nu)$, and $\Gamma(K\to \mu\nu)/\Gamma(\pi\to \mu\nu)$ are used to extract the CKM matrix element $V_{us}$. The relevant
\begin{align}\nonumber
|V^{\tau K}_{us}|\simeq&|V^{\cal L}_{us}| \Big(1+{1\over 2}\Delta_{\tau K}\Big)\;,
\\\nonumber
|V^{\tau K/\pi}_{us}|\simeq&|V^{\cal L}_{us}|  \Big(1+{1\over 2}\Delta_{\tau K}-{1\over 2}\Delta_{\tau \pi}\Big)\;,
\\
|V^{K/\pi \mu}_{us}|\simeq&|V^{\cal L}_{us}|  \Big(1+{1\over 2}\Delta_{K\mu}-{1\over 2}\Delta_{\pi\mu}\Big)\;.
\end{align}
Experimentally, the CKM factors extracted from the above processes are shown in Table~\ref{tab:CKM}. The result in each case
strongly depends on the unitarity assumption $|V_{us}^X|=\sqrt{1-|V_{ud}^X|^2-|V_{ub}|^2}$ where $X$ indicates the process used to extract $V_{ud}$.
Assuming $|V_{us}^\mathcal{L}|=0.225\pm 0.010$~\cite{Coutinho:2019aiy},
we obtain for the ratios of the measured values of $|V_{us}|$ relative to the $|V_{us}^{\mathcal{L}}|$
\begin{eqnarray}
{|V_{us}^{\tau K}|_{\rm exp}\over |V_{us}^\mathcal{L}|} = 0.993\pm { 0.045} \;,
{|V_{us}^{\tau K/\pi}|_{\rm exp}\over |V_{us}^\mathcal{L}|} = 0.994\pm {0.045} \;,
{|V_{us}^{K/\pi\mu}|_{\rm exp}\over |V_{us}^\mathcal{L}|} = 1.002\pm { 0.045}\;.
\end{eqnarray}

We finally discuss super-allowed nuclear $\beta$ decays, which probe the CKM element $|V_{ud}|$. In particular LNC vector operators with left-handed neutrinos $\nu$ interfere with the SM contribution and thus modify the extracted value $|V_{ud}|$ in $\beta$ decays, which we denote by $|V_{ud}^\beta|$. Denoting the Lagrangian value of $|V_{ud}|$ by $|V_{ud}^{\mathcal{L}}|$, we obtain in LNEFT
  \begin{align}
   \frac{|V_{ud}^\beta|}{|V_{ud}^{\mathcal{L}}|}=
\left|1 - \frac{ C_{ud\ell \nu1,\rm NP}^{V,ud ee} + C_{ud\ell\nu2}^{V,udee}
    }{2\sqrt{2} G_F V_{ud}^{\mathcal{L}}} \right|\;,
\end{align}
At the bottom of Table~\ref{tab:CKM} we quote the results for $|V_{ud}|$ extracted from $\beta$ decay in two recent analyses of Seng/Gorchtein/Ramsey-Musolf (SGR)~\cite{Seng:2018qru} and Czarnecki/Marciano/Sirlin (CMS)~\cite{Czarnecki:2019mwq}.
A comparison with the assumed Lagrangian value $|V_{ud}^{\mathcal{L}}|=\sqrt{1-(V_{us}^{\mathcal{L}})^2}=0.9744\pm0.0023$~\cite{Coutinho:2019aiy}
results in
\begin{align}
    \frac{|V_{ud}^\beta|_{\rm SGR}}{|V_{ud}^{\mathcal{L}}|}
    & =
    \frac{|V_{ud}^\beta|_{\rm CMS}}{|V_{ud}^{\mathcal{L}}|}  = 0.9996 \pm0.0024 \;.
\end{align}
The results agree due to the larger error of the assumed Lagrangian value $|V_{ud}^{\mathcal{L}}|$.

\subsection{$\beta$ decay}
\label{sec:beta}
Another constraint on charged current operators comes from $\beta$ decay~\cite{Cirigliano:2013xha,Bischer:2019ttk}.
Recently Ref.~\cite{Gonzalez-Alonso:2018omy} performed a fit to $\beta$ decay for two cases with LNC interactions: in terms of LEFT without right-handed neutrinos and LNEFT with LH neutrinos for the (axial-)vector interactions and RH neutrinos for the scalar and tensor interactions. Here we present the limits in terms of the LNEFT operator basis, which we introduced in Sec.~\ref{sec:LEFT}, and reinterpret them for the LNV case. See App.~\ref{sec:betadecay} for further details and a translation to the operator basis used in Ref.~\cite{Gonzalez-Alonso:2018omy}.

The first one is the case without RH neutrinos which is presented in Sec.~4.4 of Ref.~\cite{Gonzalez-Alonso:2018omy}. The authors  place a constraint on the LNC quark-level Wilson coefficients with LH neutrinos. In terms of the LNEFT operator basis these constraints on the Wilson coefficients evaluated at the renormalization scale $\mu=2$ GeV read
\begin{align}\nonumber
\frac{\sqrt{2}}{4 G_F V_{ud}}C_{ud\ell\nu2}^{V,udee}&=0.002\pm 0.001\pm 0.021\;, \\\nonumber
\frac{\sqrt{2}}{4G_F V_{ud}} \left(C_{ud\ell\nu1}^{S,ude\beta} + C_{ud\ell\nu2}^{S,ude\beta}\right)&=0.0014\pm 0.0020\pm 0.0003\;,\\
\frac{\sqrt{2}}{4G_F V_{ud}} C_{ud\ell\nu}^{T,ude\beta}&=-0.0007\pm 0.0012\pm 0.0001\;,
\end{align}
at 90\% CL. In our latter numerical analysis, we will translate the limits on the Wilson coefficients at the renormalization scale $\mu=2$ GeV to limits on Wilson coefficients at the scale $\Lambda_\chi$ using the RG equations in Eqs.~\eqref{RGeq}.

As the scalar and tensor contribution do not interfere with the SM contribution $C_{ud\ell\nu 1}^{V,udee}$ in contrast to the contribution of the vector operator, the results can be directly translated to constraints on LNV interactions of RH neutrinos
\begin{align}\nonumber
\frac{\sqrt{2}}{4G_F V_{ud}} \left(C_{du\ell N1}^{S,due\beta*} + C_{du\ell N2}^{S,due\beta*}\right)&=0.0014\pm 0.0020\pm 0.0003\;,\\
-\frac{\sqrt{2}}{4G_F V_{ud}} C_{du\ell N}^{T,due\beta*}&=-0.0007\pm 0.0012\pm 0.0001\;.
\end{align}

The second fit is discussed in Sec.~4.5 of Ref.\cite{Gonzalez-Alonso:2018omy} and involves LH neutrinos for the vector and axial-vector interactions and RH neutrinos for the scalar and tensor interactions. In particular, the authors placed constraints on the LNC scalar and tensor interactions of RH neutrinos $N$. In terms of the LNEFT operator basis discussed in Sec.~\ref{sec:LEFT}, the constraints read
\begin{align}
  &\left|\frac{\sqrt{2}}{4G_F V_{ud}}\left(C_{ud\ell N1}^{S,ude\beta}+C_{ud\ell N2}^{S,ude\beta}\right)\right|<0.063\;,
  &&0.006<\left|\frac{\sqrt{2}}{4G_F V_{ud}} C_{ud\ell N}^{T,ude\beta} \right|<0.024
\end{align}
at 90\% CL. Note that there is a preference for a non-vanishing tensor WC. We take a conservative approach and only take into account the upper bound, when deriving constraints.
Similarly to above, this can be translated to the case of LNV interaction of LH neutrinos $\nu$
\begin{align}
  &\left|\frac{\sqrt{2}}{4G_F V_{ud}}\left(C_{du\ell \nu1}^{S,due\beta}+C_{du\ell \nu2}^{S,due\beta}\right)\right|<0.063\;,
  &&0.006<\left|\frac{\sqrt{2}}{4G_F V_{ud}} C_{du\ell \nu}^{T,due\beta} \right|<0.024\;.
\end{align}
We provide further details and in particular the matching to the operator basis in Ref.~\cite{Gonzalez-Alonso:2018omy} in App.~\ref{sec:betadecay}.

\section{Numerical results}
\label{sec:Num}
In this section, we present the experimental constraints on the Wilson coefficients in the LNEFT and SMNEFT from the observables discussed in the previous section. We adopt the above errors for each observable and follow Eq.~(\ref{eq:bound}) to derive the constraints. In the following we assume that the mass of the sterile neutrino is negligible and one operator dominates at a time to constrain the Wilson coefficients.

\begin{table}
\centering
\resizebox{\linewidth}{!}{
\renewcommand{\arraystretch}{1.1}
\begin{tabular}{|c | l |c|c|c|c|c|}
\hline
Class & LNEFT WC &~~~decay BR~~~&
~~~LFU~~~ &~~~CKM unitarity~~~&~~~$\beta$ decay~~~& $\Lambda_{\rm LNEFT}=\left|C_i\right|^{1\over 4-d}$
\\
 &$[{\rm GeV}^{4-d}]$ &~ $\Gamma^{\pi e},~\Gamma^{\pi \mu},~\Gamma^{\tau\pi},~\Gamma^{\tau\rho}$ ~&~$R^\pi_{e/\mu},~ (g_{\tau/\mu})_\pi$~& $V_{us}^{\tau K/\pi} ,V_{us}^{ K/\pi \tau} ,V_{ud}^\beta $ &  & [GeV]
\\\hline\hline
& $C_{ud\ell\nu1,\rm NP}^{V, ude\beta}$ &$(0.89)7.6\times10^{-6}$ &\cellcolor{gray!35} $(0.049)1.8\times10^{-6}$ & $(9.0)\times10^{-8}$ & - & $(45)7.5\times 10^2$
\\\cline{2-7}
& $C_{ud\ell\nu2}^{V, ude\beta}$ &$(0.89)7.6\times10^{-6}$ &$(\fcolorbox{gray!35}{gray!35}{0.049})1.8\times10^{-6}$ & $(9.0)\times10^{-8}$ &\cellcolor{gray!35} $7.4\times 10^{-7}$ & $(4.5)1.2\times 10^3$
\\\cline{2-7}
& $C_{ud\ell\nu1(2),\rm NP}^{V, ud\mu\beta}$ & $[\fcolorbox{gray!35}{gray!35}{1.3}](0.39)5.0\times10^{-6}$ &\cellcolor{gray!35} $(0.049)1.8\times10^{-6}$ & $(1.5)9.9\times10^{-6}$  & - & $[8.9](45)7.5\times 10^2$
\\\cline{2-7}
LNC & $C_{ud\ell\nu1(2),\rm NP}^{V, ud\tau\beta}$ &$(0.34)4.7\times10^{-6}$  &\cellcolor{gray!35}$(0.22)3.8\times10^{-6}$  & $(0.17)1.0\times10^{-5}$  &- & $(21)5.1\times 10^2$
\\\cline{2-7}
$\ell\nu$ case & $C_{ud\ell\nu1(2)}^{S,ude\beta}$ &$(0.16)1.4\times10^{-9}$  &\cellcolor{gray!35} $(0.087)3.2\times10^{-10}$  & - &  $1.3\times 10^{-7}$  & $(34)5.6\times 10^4$
\\\cline{2-7}
& $C_{ud\ell\nu1(2)}^{S,ud\mu\beta}$ &$[\fcolorbox{gray!35}{gray!35}{0.47}](0.14)1.9\times10^{-7}$  &\cellcolor{gray!35} $(0.18)6.6\times10^{-8}$ & $(0.56)3.7\times10^{-7}$ & -  & $[4.6](24)3.9\times 10^3$
\\\cline{2-7}
& $C_{ud\ell\nu1(2)}^{S,ud\tau\beta}$ & $(0.21)2.9\times10^{-6}$  &\cellcolor{gray!35} $(0.14)2.4\times10^{-6}$ &$(1.0)6.5\times10^{-6}$  & -  & $(27)6.5\times 10^2$
\\\cline{2-7}
& $C_{ud\ell\nu}^{T,ude\beta}$ & - & - &  -&\cellcolor{gray!35}  $5.8\times 10^{-8}$   &$4.1\times 10^3$
\\\cline{2-7}
& $C_{ud\ell\nu}^{T,ud\tau\beta}$ &\cellcolor{gray!35} $(0.56)2.6\times 10^{-6}$ & - &  -&  -  & $(13)6.2\times 10^2$
\\\hline
&$C_{ud\ell N1(2)}^{V,ude\beta}$  & $7.6\times10^{-6}$  &\cellcolor{gray!35} $1.8\times10^{-6}$ & -  &  - & $7.5\times 10^2$
\\\cline{2-7}
&$C_{ud\ell N1(2)}^{V,ud\mu\beta}$  & $5.0\times10^{-6}$  &\cellcolor{gray!35} $1.8\times10^{-6}$  & $9.9\times10^{-6}$ & -  & $7.5\times 10^2$
\\\cline{2-7}
LNC &$C_{ud\ell N1(2)}^{V,ud\tau\beta}$  & $4.7\times10^{-6}$  &\cellcolor{gray!35} $3.8\times10^{-6}$& $1.0\times10^{-5}$ & -  & $5.1\times 10^2$
\\\cline{2-7}
$\ell N$ case &$C_{ud\ell N1(2)}^{S,ude\beta}$ & $1.4\times10^{-9}$  &\cellcolor{gray!35} $3.2\times10^{-10}$ &-  & $2.3\times 10^{-6}$   & $5.6\times 10^4$
\\\cline{2-7}
&$C_{ud\ell N1(2)}^{S,ud\mu\beta}$ & $1.9\times10^{-7}$   &\cellcolor{gray!35} $6.6\times10^{-8}$& $3.7\times10^{-7}$ & -  & $3.9\times 10^3$
\\\cline{2-7}
&$C_{ud\ell N1(2)}^{S,ud\tau\beta}$ & $2.9\times10^{-6}$ &\cellcolor{gray!35} $2.4\times10^{-6}$& $6.5\times10^{-6}$ & -  & $6.5\times 10^2$
\\\cline{2-7}
&$C_{ud\ell N}^{T,ude\beta}$ & - & - &  -&\cellcolor{gray!35} $7.3\times 10^{-7}$   & $1.1\times 10^3$
\\\cline{2-7}
& $C_{ud\ell N}^{T,ud\tau\beta}$ &\cellcolor{gray!35} $2.6\times 10^{-6}$ & - &  -& -   & $6.2\times 10^2$
\\\hline\hline 
&$C_{du\ell\nu1(2)}^{V,due\beta}$ & $7.6\times10^{-6}$ &\cellcolor{gray!35} $1.8\times10^{-6}$ & -&  -  & $7.5\times 10^2$
\\\cline{2-7}
&$C_{du\ell\nu1(2)}^{V,du\mu\beta}$ & $[2.9]5.0\times10^{-6}$ &\cellcolor{gray!35} $1.8\times10^{-6}$ &  $9.9\times10^{-6}$ & -  & $7.5\times 10^2$
\\\cline{2-7}
LNV&$C_{du\ell\nu1(2)}^{V,du\tau\beta}$ &  $4.7\times10^{-6}$  &\cellcolor{gray!35} $3.8\times10^{-6}$& $1.0\times10^{-5}$ &  - & $5.1\times 10^2$
\\\cline{2-7}
$\ell\nu$ case &$C_{du\ell\nu1(2)}^{S,due\beta}$ &$1.4\times10^{-9}$  &\cellcolor{gray!35} $3.2\times10^{-10}$ &-  & $2.3\times 10^{-6}$   &  $5.6\times 10^4$
\\\cline{2-7}
&$C_{du\ell\nu1(2)}^{S,du\mu\beta}$ & $[1.1]1.9\times10^{-7}$ &\cellcolor{gray!35} $6.6\times10^{-8}$ & $3.7\times10^{-7}$  & -  & $3.9\times 10^3$
\\\cline{2-7}
&$C_{du\ell\nu1(2)}^{S,du\tau\beta}$ & $2.9\times10^{-6}$  &\cellcolor{gray!35} $2.4\times10^{-6}$& $6.5\times10^{-6}$  & -  & $6.5\times 10^2$
\\\cline{2-7}
&$C_{du\ell\nu}^{T,due\beta}$ & - & - &  -&\cellcolor{gray!35} $7.3\times 10^{-7}$    & $1.1\times 10^3$
\\\cline{2-7}
& $C_{du\ell\nu}^{T,ud\tau\beta}$ &\cellcolor{gray!35}  $2.6\times 10^{-6}$ & - &  -&  -  & $6.2\times 10^2$
\\\hline 
&$C_{du\ell N1(2)}^{V,due\beta}$ & $7.6\times10^{-6}$  &\cellcolor{gray!35} $1.8\times10^{-6}$ & - & -  & $7.5\times 10^2$
\\\cline{2-7}
&$C_{du\ell N1(2)}^{V,du\mu\beta}$ & $5.0\times10^{-6}$  &\cellcolor{gray!35} $1.8\times10^{-6}$& $9.9\times10^{-6}$  & - & $7.5\times 10^2$
\\\cline{2-7}
LNV&$C_{du\ell N1(2)}^{V,du\tau\beta}$ & $4.7\times10^{-6}$ &\cellcolor{gray!35} $3.8\times10^{-6}$& $1.0\times10^{-5}$  & -  & $5.1\times 10^2$
\\\cline{2-7}
$\ell N$ case &$C_{du\ell N1(2)}^{S,due\beta}$ & $1.4\times10^{-9}$  &\cellcolor{gray!35} $3.2\times10^{-10}$ & -  & $1.1\times 10^{-7}$  &  $5.6\times 10^4$
\\\cline{2-7}
&$C_{du\ell N1(2)}^{S,du\mu\beta}$ & $1.9\times10^{-7}$  &\cellcolor{gray!35} $6.6\times10^{-8}$& $3.7\times10^{-7}$ & - & $3.9\times 10^3$
\\\cline{2-7}
&$C_{du\ell N1(2)}^{S,du\tau\beta}$ & $2.9\times10^{-6}$ &\cellcolor{gray!35} $2.4\times10^{-6}$& $6.5\times10^{-6}$  & -  & $6.5\times 10^2$
\\\cline{2-7}
&$C_{du\ell N}^{T,due\beta}$ & - & - &  -&\cellcolor{gray!35} $5.8\times 10^{-8}$  & $4.1\times 10^3$
\\\cline{2-7}
& $C_{du\ell N}^{T,ud\tau\beta}$ &\cellcolor{gray!35}  $2.6\times 10^{-6}$ & - &  -&  -  & $6.2\times 10^2$
\\\hline
\end{tabular}
}
\caption{Constraints on the Wilson coefficients of the dim-6 charged current operators involving a down quark in the LNEFT.
  In the LNC $\ell\nu$ case, the numbers outside (inside) of the curly bracket indicate the LFV (LFC) cases with $\beta\neq \ell$ ($\beta= \ell$). We also show the constraints for the case with $\beta=e$ from $\pi^+ \to \mu^+\bar \nu_e/\nu_e$ in Table~\ref{tab:BR} by the numbers in square brackets.}
\label{tab:LNEFTud}
\end{table}

\begin{table}
\centering
\resizebox{\linewidth}{!}{
\renewcommand{\arraystretch}{1.1}
\begin{tabular}{|c | l |c|c|c|c|}
\hline
Class &LNEFT WC &~~~decay BR~~~&
~~~LFU~~~ &~~~CKM unitarity~~~& $\Lambda_{\rm LNEFT}=\left|C_i\right|^{1\over 4-d}$
\\
& $[{\rm GeV}^{4-d}]$ &~ $\Gamma^{K e},~\Gamma^{K \mu},~\Gamma^{\tau K},~\Gamma^{\tau K^*}$ ~&~$R^K_{e/\mu}, ~(g_{\tau/\mu})_K$~& $V_{us}^{\tau K},V_{us}^{\tau K/\pi},V_{us}^{ K/\pi \tau}  $ & [GeV]
\\\hline\hline
& $C_{ud\ell\nu1(2),\rm NP}^{V, use\beta}$ &$(0.11)1.3\times10^{-6}$ &\cellcolor{gray!35} $(0.32)6.9\times10^{-7}$  & - & $(5.6)1.2\times10^3$
\\\cline{2-6}
& $C_{ud\ell\nu1(2),\rm NP}^{V, us\mu\beta}$ & $[\fcolorbox{gray!35}{gray!35}{5.4}](0.50)8.6\times10^{-7}$ &\cellcolor{gray!35}$(0.32)6.9\times10^{-7}$  &$(0.35)2.3\times10^{-6}$  & $[1.4](5.6)1.2\times10^3$
\\\cline{2-6}
LNC & $C_{ud\ell\nu1(2),\rm NP}^{V, us\tau\beta}$ &\cellcolor{gray!35}  $(0.13)1.4\times10^{-6}$ &$(0.14)1.4\times10^{-6}$ & $(0.38)2.4\times10^{-6}$ &  $(28)8.5\times10^2$
\\\cline{2-6}
$\ell\nu$ case & $C_{ud\ell\nu1(2)}^{S,use\beta}$ & $(0.22)2.5\times10^{-10}$  &\cellcolor{gray!35} $ (0.063)1.4\times10^{-10}$ & -  & $(40)8.5\times10^4$
\\\cline{2-6}
& $C_{ud\ell\nu1(2)}^{S,us\mu\beta}$ & $[\fcolorbox{gray!35}{gray!35}{2.2}](0.21)3.6\times10^{-8}$ &\cellcolor{gray!35}$(0.13)2.8\times10^{-8}$  & $(1.4)9.4\times10^{-8}$ & $[6.7](28)5.9\times10^3$
\\\cline{2-6}
& $C_{ud\ell\nu1(2)}^{S,us\tau\beta}$ &\cellcolor{gray!35}  $(0.91)9.7\times10^{-7}$  &$(0.096)1.0\times10^{-6}$ & $(0.26)1.7\times10^{-6}$ & $(3.3)1.0\times10^3$
\\\cline{2-6}
&$C_{ud\ell\nu}^{T,us\tau\beta}$ &\cellcolor{gray!35}  $(2.5)8.6\times10^{-7}$   &  - & -  &  $(2.0)1.1\times10^3$
\\\hline
&$C_{ud\ell N1(2)}^{V,use\beta}$  & $1.3\times10^{-6}$  &\cellcolor{gray!35} $6.9\times10^{-7}$  & -  & $1.2\times10^3$
\\\cline{2-6}
&$C_{ud\ell N1(2)}^{V,us\mu\beta}$  & $8.6\times10^{-7}$   &\cellcolor{gray!35} $6.9\times10^{-7}$  & $ 2.3\times10^{-6}$  & $1.2\times10^3$
\\\cline{2-6}
LNC &$C_{ud\ell N1(2)}^{V,us\tau\beta}$  &\cellcolor{gray!35}  $1.4\times10^{-6}$  & $1.4\times10^{-6}$ & $ 2.4\times10^{-6}$  & $8.5\times10^2$
\\\cline{2-6}
$\ell N$ case &$C_{ud\ell N1(2)}^{S,use\beta}$ & $2.5\times10^{-10}$  &\cellcolor{gray!35} $1.4\times10^{-10}$  & -  & $8.5\times10^4$
\\\cline{2-6}
&$C_{ud\ell N1(2)}^{S,us\mu\beta}$ &$3.6\times10^{-8}$  &\cellcolor{gray!35} $2.8\times10^{-8}$ & $ 9.4\times10^{-8}$ & $5.9\times10^3$
\\\cline{2-6}
&$C_{ud\ell N1(2)}^{S,us\tau\beta}$ &\cellcolor{gray!35}  $9.7\times10^{-7}$ &$1.0\times10^{-6}$ & $ 1.7\times10^{-6}$  & $1.0\times10^3$
\\\cline{2-6}
&$C_{ud\ell N}^{T,us\tau\beta}$ &\cellcolor{gray!35}  $8.6\times10^{-7}$   &  - & - &  $1.1\times10^3$
\\\hline\hline 
&$C_{du\ell\nu1(2)}^{V,sue\beta}$ & $1.3\times10^{-6}$  &\cellcolor{gray!35} $6.9\times10^{-7}$ & -  & $1.2\times10^3$
\\\cline{2-6}
&$C_{du\ell\nu1(2)}^{V,su\mu\beta}$ &$[\fcolorbox{gray!35}{gray!35}{5.9}]8.6\times10^{-7}$  &\cellcolor{gray!35} $6.9\times10^{-7}$  & $ 2.3\times10^{-6}$ & $[1.3]1.2\times10^3$
\\\cline{2-6}
LNV&$C_{du\ell\nu1(2)}^{V,su\tau\beta}$ &\cellcolor{gray!35} $1.4\times10^{-6}$ & $1.4\times10^{-6}$ &  $ 2.4\times10^{-6}$& $8.5\times10^2$
\\\cline{2-6}
$\ell\nu$ case &$C_{du\ell\nu1(2)}^{S,sue\beta}$ & $2.5\times10^{-10}$  &\cellcolor{gray!35} $1.4\times10^{-10}$ & -  & $8.5\times10^4$
\\\cline{2-6}
&$C_{du\ell\nu1(2)}^{S,su\mu\beta}$ &$[\fcolorbox{gray!35}{gray!35}{2.4}]3.6\times10^{-8}$ &\cellcolor{gray!35} $2.8\times10^{-8}$  & $ 9.4\times10^{-8}$  & $[6.4]5.9\times10^3$
\\\cline{2-6}
&$C_{du\ell\nu1(2)}^{S,su\tau\beta}$ &\cellcolor{gray!35} $9.7\times10^{-7}$  &$1.0\times10^{-6}$ & $ 1.7\times10^{-6}$  & $1.0\times10^3$
\\\cline{2-6}
&$C_{du\ell\nu}^{T,su\tau\beta}$ &\cellcolor{gray!35}  $8.6\times10^{-7}$   & - & -  &  $1.1\times10^3$
\\\hline 
&$C_{du\ell N1(2)}^{V,sue\beta}$ & $1.3\times10^{-6}$ &\cellcolor{gray!35} $6.9\times10^{-7}$  & -  & $1.2\times10^3$
\\\cline{2-6}
&$C_{du\ell N1(2)}^{V,su\mu\beta}$ & $8.6\times10^{-7}$  &\cellcolor{gray!35} $6.9\times10^{-7}$  & $ 2.3\times10^{-6}$ & $1.2\times10^3$
\\\cline{2-6}
LNV&$C_{du\ell N1(2)}^{V,su\tau\beta}$ &\cellcolor{gray!35}  $1.4\times10^{-6}$ & $1.4\times10^{-6}$  & $ 2.4\times10^{-6}$ & $8.5\times10^2$
\\\cline{2-6}
$\ell N$ case &$C_{du\ell N1(2)}^{S,sue\beta}$ & $2.5\times10^{-10}$  &\cellcolor{gray!35} $1.4\times10^{-10}$  & -  & $8.5\times10^4$
\\\cline{2-6}
&$C_{du\ell N1(2)}^{S,su\mu\beta}$ & $3.6\times10^{-8}$  &\cellcolor{gray!35} $2.8\times10^{-8}$ & $ 9.4\times10^{-8}$ & $5.9\times10^3$
\\\cline{2-6}
&$C_{du\ell N1(2)}^{S,su\tau\beta}$ &\cellcolor{gray!35}  $9.7\times10^{-7}$ & $1.0\times10^{-6}$  & $ 1.7\times10^{-6}$  & $1.0\times10^3$
\\\cline{2-6}
&$C_{du\ell N}^{T,su\tau\beta}$ &\cellcolor{gray!35}  $8.6\times10^{-7}$   & - & - &  $1.1\times10^3$
\\
\hline
\end{tabular}
}
\caption{
Constraints on the Wilson coefficients of the dim-6 charged current operators involving a strange quark in the LNEFT.
In the LNC $\ell\nu$ case, the numbers outside (inside) of the bracket indicate the LFV (LFC) cases with $\beta\neq\ell$ ($\beta= \ell$).  The numbers in the square bracket indicate the constraints for the case with $\beta=e$ from $\pi^+ \to \mu^+\bar \nu_e/\nu_e$ in Table~\ref{tab:BR}. In some cells, the relevant WC can be constrained by more than one observables in the class, and we only show the strongest limit obtained from the corresponding observable. For instance, in the decay BR class, $C_{ud\ell\nu1(2),\rm NP}^{us\tau\beta}$ can be constrained by  both $\Gamma^{\tau K}$ and $\Gamma^{\tau K^*}$, and we only give the strongest limit obtained from $\Gamma^{\tau K}$.}
\label{tab:LNEFTus}
\end{table}

In Tables~\ref{tab:LNEFTud} and \ref{tab:LNEFTus} we show the constraints on the Wilson coefficients of LNEFT from the
low-energy measurements in the limit of massless RH neutrinos. They correspond to the down-type quark in the 4-fermion interaction being $d$ and $s$, respectively. The last column displays the lower limit on the energy scale in LNEFT $\Lambda_{\rm LNEFT}$. In some cases the relevant WC can be constrained by more than one observable in each class, and we only show the strongest limit obtained from the corresponding observable. For instance, in the decay BR class, $C_{ud\ell\nu1(2),\rm NP}^{us\tau\beta}$ can be constrained by both $\Gamma^{\tau K}$ and $\Gamma^{\tau K^\ast}$. We only give the strongest limit obtained from $\Gamma^{\tau K}$ in Table~\ref{tab:LNEFTus}. Generally The neutrino flavor $\beta$ is arbitrary in the two tables.
In particular, the LNC $\ell\nu$ coefficients have both the LFC and LFV components and those in the LFC case have interference with the SM contribution. The numbers outside (inside) of the round brackets indicate the LFV (LFC) cases with $\beta\neq\ell$ ($\beta = \ell$). For $\beta=e$ we also show the constraints from $\pi^+ \to \mu^+\bar \nu_e/\nu_e$ in square brackets. One can see that LFU provides the most stringent constraint for the vector and scalar WCs. In addition tau lepton decay into vector meson and beta decay give sole bounds on the $\ell=\tau$ and $\ell=e$ components of tensor WCs involving a down quark, respectively. For the operators with strange quark, tau lepton decays also constrain the $\ell=\tau$ component of both scalar and tensor WCs.
The LNEFT cutoff scale for the charged current operators with down (strange) quark is at least 500 (850) GeV.

\begin{table}
\centering
\resizebox{\linewidth}{!}{
\renewcommand{\arraystretch}{1.14}
\begin{tabular}{| c | c | c | c | c | c | c | }
\hline
Class &LNEFT WC &  Limits & $\Lambda_{\rm LNEFT}=\left|C_i\right|^{1\over 4-d}$ &
LNEFT WC &  Limits  & $\Lambda_{\rm LNEFT}=\left|C_i\right|^{1\over 4-d}$
\\\hline\hline
LNC & $C_{ud\ell\nu}^{T,ude\beta}$ &\cellcolor{gray!35}  $(0.25)9.2\times10^{-9}$ &$(6.3)1.0\times 10^{4}$ &
$C_{ud\ell\nu}^{T,use\beta}$ &\cellcolor{gray!35}  $(0.18)3.9\times10^{-9}$ &  $(7.4)1.6\times 10^{4}$
\\\cline{2-7}
$\ell\nu$ case & $C_{ud\ell\nu}^{T,ud\mu\beta}$ &\cellcolor{gray!35}  $[1.3](0.052)1.9\times10^{-6}$ & $[8.6](44)7.3\times10^2 $ &
$C_{ud\ell\nu}^{T,us\mu\beta}$ &\cellcolor{gray!35}  $[6.4](0.38)8.2\times10^{-7}$  & $[1.3](5.2)1.1\times10^3$
\\\cline{2-7}
& $C_{ud\ell\nu}^{T,ud\tau\beta}$ &$(0.40)6.8\times10^{-5}$  & $(5.0)1.2\times10^2$ &
$C_{ud\ell\nu}^{T,us\tau\beta}$ & $(0.26)2.8\times10^{-5}$  & $(6.2)1.9\times10^2$
\\\hline\hline
LNC & $C_{ud\ell N}^{T,ude\beta}$ &\cellcolor{gray!35}  $9.2\times10^{-9}$ &$1.0\times 10^{4}$ &
$C_{ud\ell N}^{T,use\beta}$ &\cellcolor{gray!35}  $3.9\times10^{-9}$ &  $1.6\times 10^{4}$
\\\cline{2-7}
$\ell N$ case & $C_{ud\ell N}^{T,ud\mu\beta}$ &\cellcolor{gray!35}  $1.9\times10^{-6}$ & $7.3\times10^2 $ &
$C_{ud\ell N}^{T,us\mu\beta}$ &\cellcolor{gray!35}  $8.2\times10^{-7}$  & $1.1\times10^3$
\\\cline{2-7}
& $C_{ud\ell N}^{T,ud\tau\beta}$ & $6.8\times10^{-5}$  & $1.2\times10^2$ &
$C_{ud\ell N}^{T,us\tau\beta}$ & $2.8\times10^{-5}$  & $1.9\times10^2$
\\\hline\hline
LNV & $C_{du\ell\nu}^{T,due\beta}$ &\cellcolor{gray!35}  $9.2\times10^{-9}$ &$1.0\times 10^{4}$ &
$C_{du\ell\nu}^{T,sue\beta}$ &\cellcolor{gray!35}  $3.9\times10^{-9}$ &  $1.6\times 10^{4}$
\\\cline{2-7}
$\ell\nu$ case & $C_{du\ell\nu}^{T,du\mu\beta}$ &\cellcolor{gray!35}  $1.9\times10^{-6}$ & $7.3\times10^2 $ &
 $C_{du\ell\nu}^{T,su\mu\beta}$ &\cellcolor{gray!35}  $[7.0]8.2\times10^{-7}$ & $[1.2]1.1\times 10^3$
\\\cline{2-7}
& $C_{du\ell\nu}^{T,du\tau\beta}$ & $6.8\times10^{-5}$  & $1.2\times10^2$ &
$C_{du\ell\nu}^{T,su\tau\beta}$ & $2.8\times10^{-5}$  & $1.9\times10^2$
\\\hline\hline
LNV & $C_{du\ell N}^{T,due\beta}$ &\cellcolor{gray!35}  $9.2\times10^{-9}$ &$1.0\times 10^{4}$ &
$C_{du\ell N}^{T,sue\beta}$ &\cellcolor{gray!35}  $3.9\times10^{-9}$ &  $1.6\times 10^{4}$
\\\cline{2-7}
$\ell N$ case & $C_{du\ell N}^{T,du\mu\beta}$ &\cellcolor{gray!35}  $1.9\times10^{-6}$ & $7.3\times10^2 $ &
$C_{du\ell N}^{T,su\mu\beta}$  &\cellcolor{gray!35}  $8.2\times10^{-7}$  & $1.1\times10^3$
\\\cline{2-7}
& $C_{du\ell N}^{T,du\tau\beta}$ & $6.8\times10^{-5}$  & $1.2\times10^2$ &
$C_{du\ell N}^{T,su\tau\beta}$ & $2.8\times10^{-5}$  & $1.9\times10^2$
\\\hline
\end{tabular}
}
\caption{Constraints on the Wilson coefficients  of the dim-6 charged tensor current operators in LNEFT at the chiral symmetry breaking scale $\Lambda_{\chi}$, assuming that the tensor operator is generated at the electroweak scale $m_W$ through the RG mixing effect with the scalar current operators. We use the most stringent constraints on the Wilson coefficients of the dim-6 scalar current operators in Table~\ref{tab:LNEFTud} and Table~\ref{tab:LNEFTus} and the RG equation in Eq.~\eqref{RG:ST} to obtain the results. The notation for the square/round bracket is the same as Table~\ref{tab:LNEFTud} and Table~\ref{tab:LNEFTus}. The grey cells imply the RG mixing effect from scalar operators giving the most stringent limit for the relevant tensor Wilson coefficient compared with the direct constraint from the observables in Table~\ref{tab:LNEFTud} and Table~\ref{tab:LNEFTus}.}
\label{tab:LNEFTST}
\end{table}

As seen above, most of low-energy observables are insensitive to the tensor
operators but strongly depend on the scalar operators. Due to the RG mixing
effect between the two different types of operators in Eq.~\eqref{RG:ST}, we
can also set bounds on tensor operators from the constraints on scalar
operators. In Table~\ref{tab:LNEFTST}, we show
the constraints on the tensor WCs at the chiral symmetry breaking scale
$\Lambda_\chi$ by choosing the strongest limits on the scalar operators from
Tables~\ref{tab:LNEFTud} and \ref{tab:LNEFTus}.
These constraints assume that only a tensor operator is present at the
electroweak scale $\Lambda_{\rm EW}$, which then induces both scalar and tensor operators at the chiral symmetry breaking scale $\Lambda_{\chi}$ via RG running.
The grey cells imply that the
RG mixing effect from scalar operators gives the most stringent limit for the
relevant tensor Wilson coefficient compared with the direct constraint from
the observables in Tables~\ref{tab:LNEFTud} and \ref{tab:LNEFTus}.

Finally, we match the bounds on the LNEFT Wilson coefficients up to the SMNEFT to constrain new physics above the electroweak scale and show the results in Table~\ref{tab:SMNEFT}.
\footnote{We do not include running in the SMNEFT. See e.g. Ref.~\cite{Chala:2020pbn} for a recent discussion of 1-loop running of Higgs-neutrino operators in SMNEFT.}
The limits on new physics scale $\Lambda_{\rm NP}$ are given in units of the SM Higgs vev $v=(\sqrt{2}G_F)^{-1/2}\simeq 246$ GeV. One can see that, generally, the constraints on the NP scale for the operators with a strange quark are more stringent than those with a down quark. This is because of the enhancement by $|V_{ud}|/|V_{us}|$ in the SM contribution of pion LFU observations with respect to that of Kaon.
For the operators with down (strange) quark, the most stringent constraints on the scales of dim-6 LNC and dim-7 LNV operators are 300 (460) $v\simeq$ 74 (110) TeV and 40 (53) $v\simeq$ 9.8 (13) TeV, respectively.

\begin{table}[!h]
\centering
\resizebox{\linewidth}{!}{
\renewcommand{\arraystretch}{1.09}
\begin{tabular}{ | c | c | c | c|c|c| }
\hline
\cellcolor{gray!35} Dim-6 LNC WC & $[v^{-2}]$ & $\Lambda_{\rm NP}=|C_i|^{-\tfrac12} [v]$
&
\cellcolor{gray!35} Dim-6 LNC WC & $[v^{-2}]$ & $\Lambda_{\rm NP}=|C_i|^{-\tfrac12} [v]$
\\\hline\hline
 $|V_{xd} C_{lq}^{(3), \beta e x1 *} -V_{xd} C_{Hq}^{(3),1x} \delta^{\beta e}  -V_{ud} C_{Hl}^{(3),\beta e*}|$ &  $(0.15)5.4 \times 10^{-2}$& (26) 4.3
&$ |C_{Hud}^{11}| $ &  $2.9\times10^{-3}$ & 18
\\\hline 
$|V_{xd} C_{lq}^{(3),\beta \mu x 1 *} -V_{xd} C_{Hq}^{(3),1x} \delta^{\beta \mu}  -V_{ud} C_{Hl}^{(3),\beta \mu*}|$ &  $[3.9](0.15)5.4  \times 10^{-2}$& [5.1] (26) 4.3
& - &  - & -
 \\\hline 
$|V_{xd} C_{lq}^{(3), \beta \tau x 1 *} -V_{xd} C_{Hq}^{(3),1x} \delta^{\beta\tau}  -V_{ud} C_{Hl}^{(3),\beta \tau*}|$ &  $(0.066)1.1 \times 10^{-1}$& (12) 3.0
&- &  - & -
\\\hline 
$ |V_{xd} C_{lequ}^{(1),\beta ex1}| $ , $ |C_{ledq}^{\beta e 11}|$& $(0.03)1.1 \times 10^{-5}$& $(18)3 \times 10^2$
& $|V_{xd} C_{lequ}^{(3),\beta e x1 }|$ & $(0.18)6.7\times 10^{-4}$& (230)39
\\\hline 
$ |V_{xd} C_{lequ}^{(1),\beta\mu x1}| $ , $ |C_{ledq}^{\beta \mu 11}|$& $[1.6](0.061)2.2 \times 10^{-3}$&[25](130)21
& $|V_{xd} C_{lequ}^{(3),\beta\mu x1 }|$ &$[0.094](0.0038)0.14$ & $[3.3](16)2.7$
\\\hline 
%
$| V_{xd} C_{lequ}^{(1),\beta\tau x1}| $ , $ |C_{ledq}^{\beta \tau 11 }|$& $(0.48)8.2 \times 10^{-2}$ &  (14)3.5
& $|V_{xd} C_{lequ}^{(3),\beta\tau x1 }|$ &$(0.041)0.19$& (5.0) 2.3
\\\hline\hline
 $|V_{ud} C_{HNe}^{\beta e}|$&  $0.11$ & 3.0
&$ |C_{duNe}^{11\beta e} | $&  $0.11 $ & 3.0
\\\hline 
$|V_{ud} C_{HNe}^{\beta \mu}|$&  $0.11$& 3.0
&$ |C_{duNe}^{11\beta \mu} | $&  $0.11$ & 3.0
\\\hline 
$|V_{ud} C_{HNe}^{\beta \tau}|$&  $0.23$& 2.1
&$ |C_{duNe}^{11\beta \tau} | $&  $0.23$ & 2.1
\\\hline 
$ |C_{LNQd}^{e\beta 11}-\tfrac12 C_{LdQN}^{e 11 \beta}| $, $|V_{xd} C_{QuNL}^{x1\beta e*}|$ & $1.1 \times 10^{-5}$& $300$
& $|C_{LdQN}^{e11\beta}|$ & $5.3\times 10^{-3}$&  14
\\\hline 
$ |C_{LNQd}^{\mu\beta 11}-\tfrac12 C_{LdQN}^{\mu 11 \beta}| $, $|V_{xd} C_{QuNL}^{x1\beta \mu*}|$ & $2.2 \times 10^{-3}$& $21$
& $|C_{LdQN}^{\mu 11\beta}|$ & $1.1$&  0.95
\\\hline 
$ |C_{LNQd}^{\tau\beta 11}-\tfrac12 C_{LdQN}^{\tau 11 \beta}| $, $|V_{xd} C_{QuNL}^{x1\beta \tau*}|$ & $8.2\times 10^{-2}$& $3.5$
& $|C_{LdQN}^{\tau 11\beta}|$ & $1.5$&  0.81
\\\hline 
\end{tabular}
}



\resizebox{\linewidth}{!}{
\renewcommand{\arraystretch}{1.09}

\begin{tabular}{|c|c|c|c|c|c|}
\hline
\cellcolor{gray!35} Dim-7 LNV WC & $[v^{-3}]$ & $\Lambda_{\rm NP}=|C_i|^{-\tfrac13} [v]$
&
\cellcolor{gray!35} Dim-7 LNV WC & $[v^{-3}]$ & $\Lambda_{\rm NP}=|C_i|^{-\tfrac13} [v]$
\\\hline\hline

  $|V_{ud} C_{LeHD}^{\beta e}|$, $|V_{xd}^* C_{QNLH2}^{x1\beta e} +V_{ud} C_{NL1}^{\beta e}|$ &  $0.15$& 1.9
& $|C_{\bar duLeH}^{ 11\beta e}|$, $|C_{duNLH}^{11\beta e}|$ &  $0.15 $ & 1.9
\\\hline 
  $|V_{ud} C_{LeHD}^{\beta\mu}|$, $|V_{xd}^* C_{QNLH2}^{x1\beta \mu} + V_{ud} C_{NL1}^{\beta \mu}|$ &  $0.15$& 1.9
& $|C_{\bar duLeH}^{11\beta\mu}|$, $|C_{duNLH}^{11\beta \mu}|$ &  $0.15$ & 1.9
\\\hline 
  $|V_{ud} C_{LeHD}^{\beta \tau}|$, $|V_{xd}^* C_{QNLH2}^{x1\beta\tau} + V_{ud} C_{NL1}^{\beta \tau}|$ &  $0.32$& 1.5
& $|C_{\bar duLeH}^{11\beta\tau}|$, $|C_{duNLH}^{11\beta\tau}|$ &  $0.32$ & 1.5
\\\hline 
$|C_{\bar d Q LLH1}^{11e\beta}|$, $|V_{xd}^*C_{\bar Q uLLH}^{x1e\beta}|$, $|V_{xd}^* C_{QuNeH1}^{x1\beta e}|$, $|C_{dQNeH}^{11\beta e}|$ & $1.5 \times 10^{-5}$& $40$
& $|C_{\bar d QLLH2}^{11e\beta}|$, $|V_{xd}^* C_{QuNeH2}^{x1\beta e}|$ & $9.4\times10^{-4}$&  10
\\\hline 
$|C_{\bar d Q LLH1}^{11\mu\beta}|$, $|V_{xd}^*C_{\bar Q uLLH}^{x1 \mu\beta}|$, $|V_{xd}^* C_{QuNeH1}^{x1\beta \mu}|$, $|C_{dQNeH}^{11\beta \mu}|$ & $3.2 \times 10^{-3}$& $6.8$
& $|C_{\bar d QLLH2}^{11\mu\beta}|$, $|V_{xd}^* C_{QuNeH2}^{x1\beta \mu}|$ & $0.20$&  1.7
\\\hline 
$|C_{\bar d Q LLH1}^{11\tau\beta}|$, $|V_{xd}^*C_{\bar Q uLLH}^{x1\tau\beta}|$, $|V_{xd}^* C_{QuNeH1}^{x1\beta \tau}|$, $|C_{dQNeH}^{11\beta \tau}|$ & $0.12$& $2.1$
& $|C_{\bar d QLLH2}^{11\tau\beta}|$, $|V_{xd}^* C_{QuNeH2}^{x1\beta \tau}|$ & $0.27$&  1.6
\\\hline

\end{tabular}
}


\vspace{3ex}

\resizebox{\linewidth}{!}{
\renewcommand{\arraystretch}{1.2}
\begin{tabular}{ | c | c | c | c|c|c| }
\hline
\cellcolor{gray!35} Dim-6 LNC WC & $[v^{-2}]$ & $\Lambda_{\rm NP}=|C_i|^{-\tfrac12} [v]$
&
\cellcolor{gray!35} Dim-6 LNC WC & $[v^{-2}]$ & $\Lambda_{\rm NP}=|C_i|^{-\tfrac12} [v]$
\\\hline\hline
 $|V_{xs} C_{lq}^{(3), \beta e x 1 *} -V_{xs} C_{Hq}^{(3),1x} \delta^{\beta e}  -V_{us} C_{Hl}^{(3),\beta e*}|$ &  $(0.096)2.1\times 10^{-2}$& (32) 6.9
&$ |C_{Hud}^{12}| $ &  $1.9\times10^{-3} $ & 23
\\\hline 
$|V_{xs} C_{lq}^{(3), \beta \mu x 1 *} -V_{xs} C_{Hq}^{(3),1x} \delta^{\beta\mu}  -V_{us} C_{Hl}^{(3),\beta \mu*}|$ &  $[1.6](0.096)2.1  \times 10^{-2}$& [7.9] (32) 6.9
& - &  -& -
 \\\hline 
$|V_{xs} C_{lq}^{(3), \beta \tau x 1 *} -V_{xs} C_{Hq}^{(3),1x} \delta^{\beta\tau}  -V_{us} C_{Hl}^{(3),\beta \tau*}|$ &  $(0.39)4.2 \times 10^{-2}$& (16) 4.9
& - &  - & -
\\\hline 
$ |V_{xs} C_{lequ}^{(1),\beta e x1}| $ , $ |C_{ledq}^{\beta e 21}|$& $(0.21)4.8 \times 10^{-6}$& $(22)4.6\times 10^2$
& $|V_{xs} C_{lequ}^{(3),\beta e x1 }|$ & $(0.13)2.8\times 10^{-4} $&  (280) 59
\\\hline 
$ |V_{xs} C_{lequ}^{(1),\beta\mu x1}| $ , $ |C_{ledq}^{\beta\mu 21 }|$& $[7.5](0.44)9.5 \times 10^{-4}$& [37](150)32
& $|V_{xs} C_{lequ}^{(3),\beta\mu x1 }|$ &$[4.6](0.28)6.0\times10^{-2}$&$[4.6](19)4.1$
\\\hline 
%
$| V_{xs} C_{lequ}^{(1),\beta\tau x1}| $ , $ |C_{ledq}^{\beta\tau 21 }|$& $(0.31)3.3 \times 10^{-2}$ &  (18)5.5
& $|V_{xs} C_{lequ}^{(3),\beta\tau x1 }|$ & $(1.8)6.2\times 10^{-2}$&(7.4)4.0
\\\hline\hline
 $|V_{us} C_{HNe}^{\beta e}|$&  $0.041$ & 4.9
&$ |C_{duNe}^{21\beta e} | $&  $0.041$ & 4.9
\\\hline 
$|V_{us} C_{HNe}^{\beta \mu}|$&  $0.041$& 4.9
&$ |C_{duNe}^{21\beta \mu} | $&  $0.041$ & 4.9
\\\hline 
$|V_{us} C_{HNe}^{\beta \tau}|$&  $0.084$& 3.4
&$ |C_{duNe}^{21\beta \tau} | $&  $0.084$ & 3.4
\\\hline 
$ |C_{LNQd}^{e\beta 12}-\tfrac12 C_{LdQN}^{e 21 \beta}| $, $|V_{xs} C_{QuNL}^{x1\beta e*}|$ & $4.8 \times 10^{-6}$& $460$
& $|C_{LdQN}^{e21\beta}|$ & $0.0023$&  21
\\\hline 
$ |C_{LNQd}^{\mu\beta 12}-\tfrac12 C_{LdQN}^{\mu 21 \beta}| $, $|V_{xs} C_{QuNL}^{x1\beta \mu*}|$ & $9.5 \times 10^{-4}$& $32$
& $|C_{LdQN}^{\mu 21\beta}|$ & $0.48$&  1.4
\\\hline 
$ |C_{LNQd}^{\tau\beta 12}-\tfrac12 C_{LdQN}^{\tau 21 \beta}| $, $|V_{xs} C_{QuNL}^{x1\beta \tau*}|$ & $3.3\times 10^{-2}$& $5.5$
& $|C_{LdQN}^{\tau 21\beta}|$ & $0.50$&  1.4
\\\hline 
\end{tabular}
}



\resizebox{\linewidth}{!}{
\renewcommand{\arraystretch}{1.14}

\begin{tabular}{|c|c|c|c|c|c|}
\hline
\cellcolor{gray!35} Dim-7 LNV WC & $[v^{-3}]$ & $\Lambda_{\rm NP}=|C_i|^{-\tfrac13} [v]$
&
\cellcolor{gray!35} Dim-7 LNV WC & $[v^{-3}]$ & $\Lambda_{\rm NP}=|C_i|^{-\tfrac13} [v]$
\\\hline\hline

  $|V_{us} C_{LeHD}^{\beta e}|$, $|V_{xs}^* C_{QNLH2}^{x1\beta e} + V_{us} C_{NL1}^{\beta e}|$ &  $5.9\times 10^{-2}$& 2.6
& $|C_{\bar duLeH}^{ 21\beta e}|$, $|C_{duNLH}^{21\beta e}|$ &  $5.9\times 10^{-2} $ & 2.6
\\\hline 
  $|V_{us} C_{LeHD}^{\beta\mu}|$, $|V_{xs}^* C_{QNLH2}^{x1\beta \mu} + V_{us} C_{NL1}^{\beta \mu}|$ &  $[5.0]5.9\times 10^{-2} $& [2.7]2.6
& $ |C_{\bar duLeH}^{ 21\beta\mu}|$, $|C_{duNLH}^{21\beta \mu}|$ &  $[5.0]5.9\times 10^{-2}$ & [2.7]2.6
\\\hline 
  $|V_{us} C_{LeHD}^{\beta \tau}|$, $|V_{xs}^* C_{QNLH2}^{x1\beta\tau} + V_{us} C_{NL1}^{\beta \tau}|$ &  $0.12$& 2.0
& $|C_{d\bar uLeH}^{21\beta\tau}|$, $|C_{duNLH}^{21\beta\tau}|$ &  $0.12$ & 2.0
\\\hline 
$|C_{\bar d Q LLH1}^{21e\beta}|$, $|V_{xs}^*C_{\bar Q uLLH}^{x1 e\beta}|$, $|V_{xs}^* C_{QuNeH1}^{x1\beta e}|$, $|C_{dQNeH}^{21\beta e}|$ & $6.7 \times 10^{-6}$& $53$
& $|C_{\bar d QLLH2}^{21e\beta}|$, $|V_{xs}^* C_{QuNeH2}^{x1\beta e}|$ & $4.0\times10^{-4}$&  14
\\\hline 
$|C_{\bar d Q LLH1}^{21\mu\beta}|$, $|V_{xs}^*C_{\bar Q uLLH}^{x1 \mu\beta}|$, $|V_{xs}^* C_{QuNeH1}^{x1\beta \mu}|$, $|C_{dQNeH}^{21\beta \mu}|$ & $[1.2]1.3 \times 10^{-3}$& $[9.5]9.1$
& $|C_{\bar d QLLH2}^{21\mu\beta}|$, $|V_{xs}^* C_{QuNeH2}^{x1\beta \mu}|$ & $[7.2]8.4\times10^{-2}$&  [2.4]2.3
\\\hline 
$|C_{\bar d Q LLH1}^{21\tau\beta}|$, $|V_{xs}^*C_{\bar Q uLLH}^{x1 \tau\beta}|$, $|V_{xs}^* C_{QuNeH1}^{x1\beta \tau}|$, $|C_{dQNeH}^{21\beta \tau}|$ & $4.7\times 10^{-2}$& $2.8$
& $|C_{\bar d QLLH2}^{21\tau\beta}|$, $|V_{xs}^* C_{QuNeH2}^{x1\beta \tau}|$ & $8.8\times10^{-2}$&  2.2
\\\hline
\end{tabular}
}
\caption{Constraints on SMNEFT operators at the electroweak scale. The top (bottom) half of the table lists constraints for operators with a down (strange) quark. The notation for the brackets follows Tables~\ref{tab:LNEFTud} and \ref{tab:LNEFTus}. The generation of the quark fields on the Wilson coefficient is labeled by the number 1, 2, 3 with the identification for the CKM elements: $V_{1d(s)}\equiv V_{ud(s)}, V_{2d(s)}\equiv V_{cd(s)},V_{3d(s)}\equiv V_{td(s)}$.
}
\label{tab:SMNEFT}
\end{table}

The constraints at the electroweak scale can be related to other scales using the renormalization group equations in SMNEFT~\cite{Jenkins:2013zja,Jenkins:2013wua,Alonso:2013hga,Liao:2019tep}. Running from the electroweak scale to a scale of $\mathcal{O}(10^3~{\rm TeV})$, the dominant contribution comes from QCD corrections which only contribute to scalar and tensor quark current operators at 1-loop order. It however only leads to a minor correction. For example the Fig.~2 in Ref.~\cite{Liao:2019tep} shows that the corrections to scale of first-generation Wilson coefficients are usually below 10-20\%. The only exception is the operator $\mathcal{O}_{\bar d LQLH2}$ which is related to the tensor operator $\mathcal{O}_{\bar dQLLH2}$ and receives a correction of order 50\% due to the combined effect of strong interactions and the large top quark Yukawa coupling. Moreover, the operator mixing induced by
Yukawa couplings and electroweak corrections is negligible for light quarks, because it is suppressed by small Yukawa couplings and electroweak gauge parameters.

\section{Other constraints and vector meson decays}
\label{sec:Other}

The dominant decay channels for light vector mesons are 2-body hadronic processes mediated by strong interactions, e.g. $\rho\to \pi\pi$ and $K^{\ast}\to K\pi$. SM weak decays are highly suppressed by the $W$ boson exchange and have not been observed yet in low-energy experiments. Any excess events would indicate the existence of NP beyond the SM.
The general LNC (LNV) partial decay widths of vector mesons
$V^+\equiv V^+(u_p \bar d_r)$
to a charged lepton $\ell^+$ and a neutrino $X$ (anti-neutrino $\bar X$) with $X\in\{\nu, N\}$
are
\begin{eqnarray}
\Gamma(V^+\to \ell^+_\alpha X_\beta) &=& {m_V^3\over 48\pi}  \Big[{1\over 2}|C^V_{ud\ell X 1}+C^V_{ud\ell X 2}|^2 f_V^2 + 4 |C^T_{ud\ell X}|^2 (f_V^{T})^2\Big]\;,\\
\Gamma(V^+\to \ell^+_\alpha \bar{X}_\beta) &=& {m_V^3\over 48\pi} \Big[{1\over 2}|C^V_{du\ell X 1}+C^V_{du\ell X 2}|^2 f_V^2 + 4 |C^T_{du\ell X}|^2 (f_V^{T})^2\Big]\;,
\end{eqnarray}
in the limit of massless charged leptons and neutrinos. Given the SM part in $C^V_{ud\ell\nu 1}$, the SM prediction for vector meson decay is
$\Gamma(V^+\to \ell^+_\alpha \nu_\alpha)_{\rm SM} = {G_F^2 |V_{pr}|^2\over 12\pi} f_V^2 m_V^3$,
and thus the branching ratios for the leptonic decay of $\rho$ and $K^\ast$ within the SM are
${\cal B}(\rho^+\to \ell^+\nu_\ell) \approx 4.5\times 10^{-13}$ and  ${\cal B}(K^{\ast +}\to \ell^+\nu_\ell) \approx 1.4\times 10^{-13}$.
Taking into account the discussed constraints on the NP vector and tensor WCs, we find that the allowed branching ratio for any NP contribution to the leptonic decay of vector mesons is at least two orders of magnitude smaller than the SM predictions. Hence, it is unlikely to observe the NP through the weak decays of vector mesons.

The dim-7 SMNEFT operators matched in Table~\ref{matching} can also contribute to the neutrinoless double beta decay ($0\nu\beta\beta$) through the long distance mechanism mediated by light neutrinos~\cite{Bolton:2020xsm}. The authors of Refs.~\cite{Cirigliano:2017djv,Liao:2019tep} considered the constraints on the dim-7 SMEFT operators without RH neutrinos from the $0\nu\beta\beta$. The most stringent bounds are set by the experimental limit on the half-life of $^{136}{\rm Xe}$~\cite{KamLAND-Zen:2016pfg}, which translate into a lower bound on the NP scale in our operator convention as~\cite{Cirigliano:2017djv,Liao:2019tep}
\begin{eqnarray}
\left(C_{\bar d uLeH}^{11ee}\right)^{-1/3}\gtrsim{\cal O}(10~{\rm TeV})\;,
\qquad
\left\{C_{LeHD}^{ee},C_{\bar QuLLH}^{11ee},C_{\bar dLQLH1(2)}^{11ee} \right \}^{-1/3}\gtrsim{\cal O}(100~{\rm TeV)}\;,~
\end{eqnarray}
Compared with the low energy observables studied above, we see the $0\nu\beta\beta$ puts the most stringent constraints on the $ee$ lepton flavors.  Such results imply the $0\nu\beta\beta$ could also constrain the dim-7 SMNEFT operators involving RH neutrinos with a similar precision~\cite{Dekens:2020ttz}. We leave a detailed study on the $0\nu\beta\beta$ from the dim-7 SMNEFT operators involving RH neutrinos in Table~\ref{tab:SMNEFT7} for future work.

Recently, the constraints on SMNEFT WCs from searches at the LHC and HERA and sensitivities at future colliders have been studied in Refs.~\cite{Ruiz:2017nip,Cai:2017mow,Alcaide:2019pnf,Han:2020pff,Biekotter:2020tbd}. In particular, the authors of Ref.~\cite{Han:2020pff} studied scalar and tensor operators with sterile neutrinos and recast the recent ATLAS search~\cite{Aad:2019wvl} for a charged lepton and missing transverse momentum events.
Translated to our operator basis, the constraints on the scalar operators read
\begin{align}\nonumber
    |C_{QuNL}^{11ee}| & \leq 5.0 (0.88)\times 10^{-3} v^{-2}\;, &
    |C_{QuNL}^{11\mu\mu}| & \leq 5.8 (1.3)\times 10^{-3} v^{-2}\;, \\\nonumber
    |C_{LNQd}^{ee11}| & \leq 5.2 (0.92)\times 10^{-3} v^{-2}\;, &
    |C_{LNQd}^{\mu\mu11}| & \leq 6.0(1.4)\times 10^{-3} v^{-2}\;,\\
    |C_{LdQN}^{e11e}| & \leq 1.9 (0.38)\times 10^{-2} v^{-2}\;, &
    |C_{LdQN}^{\mu11\mu}| & \leq 2.2 (0.64)\times 10^{-2} v^{-2}\;,
    \label{eq:LHC}
\end{align}
where the constraints outside of the brackets only use the transverse mass ($m_T$) distribution below $800$ GeV, in order to ensure the validity of the effective field theory, while the constraints inside the brackets use the full $m_T$ distribution to place constraints. The strongest (weakest) constraint is on $|C_{QuNL}^{11ee}|$ ($|C_{LdQN}^{\mu 11\mu}|$) and corresponds to a lower bound on the NP scale $\Lambda_{\rm NP} \geq 25 (140)$ TeV ($\Lambda_{\rm NP}\geq 5.5 (19)$ TeV). See Ref.~\cite{Han:2020pff} for further details on current collider constraints and sensitivities of future collider searches.
As the final-state neutrino is not detected, these constraints more generally apply to arbitrary final-state neutrinos and antineutrinos. Hence the constraints in Eq.~\eqref{eq:LHC} equally apply to a general neutrino flavour $\beta$, which is not necessarily equal to the flavour of the charged lepton. Moreover, the constraints can be translated to LNV operators with LH neutrinos. We find the following constraints for the dim-7 LNV SMNEFT operators
\begin{align}\nonumber
    |C_{\bar Q uLLH}^{11ee}| & \leq 7.1 (1.2)\times 10^{-3} v^{-3}\;, &
    |C_{\bar Qu LLH}^{11\mu\mu}| & \leq 8.2 (1.9)\times 10^{-3} v^{-3}\;, \\\nonumber
    |C_{\bar d Q LLH1}^{11ee}| & \leq 7.4 (1.3)\times 10^{-3} v^{-3}\;, &
    |C_{\bar d Q LLH1}^{11\mu\mu}| & \leq 8.5 (1.9)\times 10^{-3} v^{-3}\;,\\
    |C_{\bar d Q LLH2}^{11ee}| & \leq 3.4 (0.68)\times 10^{-3} v^{-3}\;, &
    |C_{\bar d Q LLH2}^{11\mu\mu}| & \leq 4.0 (1.1)\times 10^{-3} v^{-3}\;.
\end{align}
The tensor operator has to satisfy the most stringent constraint. The strongest (weakest) constraint is on $|C_{\bar d QLLH2}^{11ee}|$ ($|C_{\bar d QLLH1}^{11\mu\mu}$) and correponds to a lower bound on the NP scale $\Lambda_{\rm NP}\geq 1.6 (2.8)$ TeV ($\Lambda_{\rm NP}\geq 1.2(2.0)$ TeV). From Table~\ref{tab:SMNEFT} we see the low energy observables place more stringent constraints on the NP scale for the relevant Wilson coefficient than the high energy LHC observables.

\section{Conclusions}
\label{sec:Con}

We investigate the constraints from low-energy precision measurements on the charged currents in general neutrino interactions with RH neutrinos in effective field theories. The interactions between the charged lepton, quarks and left-handed SM neutrinos and/or right-handed neutrinos are first described by the LNEFT between the electroweak scale and the chiral symmetry breaking scale.  We consider the most sensitive low-energy probes from the weak leptonic decays of pseudoscalar mesons and hadronic tau lepton decays. The experimental observables include precise decay branching ratios, the lepton flavor universality in pseudoscalar meson and tau lepton decays and the CKM unitarity. We also take into account the constraint on charged currents in LNEFT from nuclear beta decay and predict the weak decays of light vector mesons. Finally, we include the one-loop QCD/QED running for the LNEFT Wilson coefficients from chiral symmetry breaking scale to the electroweak scale. The bounds on the LNEFT Wilson coefficients are then matched up to the SMNEFT to constrain new physics above the electroweak scale.

We summarize our main conclusions in the following
\begin{itemize}
  \item The LFU in pseudoscalar meson and tau lepton decays provides the most stringent constraint on the vector and scalar WCs. The tau lepton decay into a vector meson and a neutrino as well as nuclear beta decay also provide complementary bounds on the tensor WCs with $\tau$ or $e$ charged lepton flavor. The LNEFT cutoff scale for the charged current operators with down (strange) quark is at least 500 (850) GeV.
\item The RG mixing effect between the scalar and tensor types of operators leads to additional constraints on the tensor WCs based on the strong limits on the scalar operators from low-energy measurements. For $\ell=e (\tau)$ tensor WCs, the RG mixing effect induces more (less) severe constraints than the beta (tau lepton) decay. It also provides complementary bounds on the $\ell=\mu$ tensor WCs.
\item The constraints on the vector and tensor WCs in LNEFT set upper limits on the weak decay branching fractions of light vector mesons. The allowed limit for leptonic decay of vector mesons are at least two orders of magnitude lower than the SM predictions.
\item The most stringent bounds on the NP scale of dim-6 LNC and dim-7 LNV operators in SMNEFT are 74 (110) TeV and 9.8 (13) TeV, respectively, for the operators with down (strange) quark.
\end{itemize}

\section*{ACKNOWLEDGMENTS}
TL would like to thank Xue-Qian Li and Mao-Zhi Yang for very useful discussion and communication. MS would like to thank Yi Cai for useful discussions.
TL is supported by the National Natural Science Foundation of China (Grant No. 11975129, 12035008) and ``the Fundamental Research Funds for the Central Universities'', Nankai University (Grant No. 63196013). XDM is supported by the MOST (Grants No. MOST 109-2811-M-002-535, No. MOST 106-2112-M-002-003-MY3). MS acknowledges support by the Australian Research Council via the Discovery Project DP200101470.

\appendix

\section{The complete operator basis involving RH neutrinos $N$ in the LNEFT}
\label{sec:LNEFTbasis}
For the convenience of the reader we make the connection to the LEFT operator basis in Ref.~\cite{Jenkins:2017jig}, which is defined as
\begin{equation}
 \mathcal{L} = \sum_i L_i \mathcal{O}_i
\end{equation}
with operators $\mathcal{O}_i$ as defined in
\begin{align}\nonumber
\calO_{\nu \ell du}^{V,LL}=&(\overline{\nu}\gamma^\mu \ell_L)(\overline{d_L}\gamma_\mu u_L)\;,
&
\calO_{\nu \ell du}^{V,LR}=&(\overline{\nu}\gamma^\mu \ell_L)(\overline{d_R}\gamma_\mu u_R)\;,
&
\calO_{\nu \ell du}^{S,RR}=&(\overline{\nu} \ell_R)(\overline{d_L}u_R)\;,
\\\nonumber
\calO_{\nu \ell du}^{S,RL}=&(\overline{\nu} \ell_R)(\overline{d_R}u_L)\;,
&
\calO_{\nu \ell du}^{T,RR}=&(\overline{\nu}\sigma^{\mu\nu} \ell_R)(\overline{d_L}\sigma_{\mu\nu}u_R)\;,
\\\nonumber
\calO_{\nu \ell du}^{V,RL}=&(\overline{\nu^C}\gamma^\mu \ell_R)(\overline{d_L}\gamma_\mu u_L)\;,
&
\calO_{\nu \ell du}^{V,RR}=&(\overline{\nu^C}\gamma^\mu \ell_R)(\overline{d_R}\gamma_\mu u_R)\;,
&
\calO_{\nu \ell du}^{S,LL}=&(\overline{\nu^C} \ell_L)(\overline{d_R}u_L)\;,
\\
\calO_{\nu \ell du}^{S,LR}=&(\overline{\nu^C} \ell_L)(\overline{d_L}u_R)\;,
&
\calO_{\nu \ell du}^{T,LL}=&(\overline{\nu^C}\sigma^{\mu\nu} \ell_L)(\overline{d_R}\sigma_{\mu\nu}u_L)\;.
\end{align}
The mapping of the subset of the operators in our basis, which maps to the operators above is given by
  \begin{align}\nonumber
    C^V_{ud\ell\nu1} & = \left(L^{V,LL}_{\nu \ell du}\right)^* &
    C^V_{ud\ell\nu2} & = \left(L^{V,LR}_{\nu \ell du}\right)^* &
    C^S_{ud\ell\nu1} & = \left(L^{S,RR}_{\nu \ell du}\right)^* \\\nonumber
    C^S_{ud\ell\nu2} & = \left(L^{S,RL}_{\nu \ell du}\right)^* &
    C^T_{ud\ell\nu} & = \left(L^{T,RR}_{\nu \ell du}\right)^* \\\nonumber
    C^V_{du\ell\nu1} & = L^{V,RL}_{\nu \ell du} &
    C^V_{du\ell\nu2} & = L^{V,RR}_{\nu \ell du} &
    C^S_{du\ell\nu1} & = L^{S,LL}_{\nu \ell du} \\
    C^S_{du\ell\nu2} & = L^{S,LR}_{\nu \ell du} &
    C^T_{du\ell\nu} & = -L^{T,LL}_{\nu \ell du} &
  \end{align}

The full list of independent LNEFT operators with at least one RH neutrino $N$ at dim-6 was constructed in Ref.~\cite{Chala:2020vqp,Li:2020lba} and is repeated in Tables~\ref{tab:LNEFTA} and \ref{tab:LNEFTB}, where in the third and sixth columns in each table we also show the independent number of operators with flavors being considered (the sterile neutrinos with $n_s$ flavors, the up-type quarks with the total flavors $n_u=2$, the remaining fermions have $n_f=3$ flavors). All those operators are classified in terms of the net number of the SM global baryon and lepton quantum numbers.

\begin{table}[!h]
\footnotesize
\center
\resizebox{\linewidth}{!}{
	\renewcommand{\arraystretch}{0.9}
\begin{tabular}{|l | l |  l|l | l | l|}
\hline
 Operator & Specific form & $\# (n_f,~n_u,~n_s)$ &  Operator & Specific form & $\# (n_f,~n_u,~n_s)$
\\
\hline
\hline
 \multicolumn{6}{|c|}{$(\Delta L, \Delta B)=(0,~0)$}
\\
\hline
\multicolumn{3}{|c|}{\cellcolor{gray!35}$(\overline{L}L)(\overline{R}R)$}
&
\multicolumn{3}{|c|}{\cellcolor{gray!35}$(\overline{R}R)(\overline{R}R)$}
\\
\hline
$\calO_{\ell N1}^V(\star\star)(\text{H})$ & $(\overline{\ell_L}\gamma^\mu \ell_L)(\overline{N}\gamma_\mu N)$ & $n_f^2n_s^2 $ &
$\calO_{\ell N2}^V(\star\star)(\text{H})$ & $(\overline{\ell_R}\gamma^\mu \ell_R)(\overline{N}\gamma_\mu N)$ & $n_f^2n_s^2$
\\
 $\calO_{dN1}^V(\star\star)(\text{H})$ & $(\overline{d_L}\gamma^\mu d_L)(\overline{N}\gamma_\mu N)$ & $n_f^2n_s^2 $ &
 $\calO_{dN2}^V(\star\star)(\text{H})$ & $(\overline{d_R}\gamma^\mu d_R)(\overline{N}\gamma_\mu N)$ & $n_f^2n_s^2 $
\\
$\calO_{uN1}^V(\star\star)(\text{H})$ & $(\overline{u_L}\gamma^\mu u_L)(\overline{N}\gamma_\mu N)$ & $n_u^2n_s^2  $ &
$\calO_{uN2}^V(\star\star)(\text{H})$ & $(\overline{u_R}\gamma^\mu u_R)(\overline{N}\gamma_\mu N)$ & $n_u^2n_s^2  $
\\
$\calO_{ud\ell N1}^V(\star)$ & $(\overline{u_L}\gamma^\mu d_L)(\overline{\ell_R}\gamma_\mu N)$ & $n_f^2n_un_s $ &
$\calO_{ud\ell N2}^V(\star)$ & $(\overline{u_R}\gamma^\mu d_R)(\overline{\ell_R}\gamma_\mu N)$ & $n_f^2n_un_s $
\\
$\calO_{\nu N}^V(\star\star)(\text{H})$ & $(\overline{\nu}\gamma^\mu \nu)(\overline{N}\gamma_\mu N)$ &
$n_f^2n_s^2 $ &
$\calO_{N}^V(\star\star\star\star)(\text{H})$ & $(\overline{N}\gamma^\mu N)(\overline{N}\gamma_\mu N)$ &
$\frac{1}{4}n_s^2(n_s+1)^2 $
\\
\hline
\multicolumn{3}{|c|}{\cellcolor{gray!35}$(\overline{L}R)(\overline{L}R)$}
&
\multicolumn{3}{|c|}{\cellcolor{gray!35}$(\overline{R}L)(\overline{L}R)$}
\\
\hline
$\calO_{\ell \nu N1}^S(\star)$ & $(\overline{\ell_L}\ell_R)(\overline{\nu}N)$ & $n_f^3n_s $ &
$\calO_{\ell \nu N2}^S(\star)$ & $(\overline{\ell_R}\ell_L)(\overline{\nu}N)$ & $n_f^3n_s $
\\
$\calO_{\ell \nu N}^T(\star)$ & $(\overline{\ell_L}\sigma^{\mu\nu}\ell_R)(\overline{\nu}\sigma_{\mu\nu}N)$ & $n_f^3n_s $ &
& &
\\
$\calO_{d\nu N1}^S(\star)$ & $(\overline{d_L}d_R)(\overline{\nu}N)$ & $n_f^3n_s $ &
$\calO_{d\nu N2}^S(\star)$ & $(\overline{d_R}d_L)(\overline{\nu}N)$ & $n_f^3n_s $
\\
$\calO_{d\nu N}^T(\star)$ & $(\overline{d_L}\sigma^{\mu\nu}d_R)(\overline{\nu}\sigma_{\mu\nu}N)$ & $n_f^3n_s $ &  & &
\\
$\calO_{u\nu N1}^S(\star)$ & $(\overline{u_L}u_R)(\overline{\nu}N)$ & $n_fn_u^2n_s$ &
$\calO_{u\nu N2}^S(\star)$ & $(\overline{u_R}u_L)(\overline{\nu}N)$ & $n_fn_u^2n_s$
\\
 $\calO_{u\nu N}^T(\star)$ & $(\overline{u_L}\sigma^{\mu\nu}u_R)(\overline{\nu}\sigma_{\mu\nu}N)$ & $n_fn_u^2n_s  $ &  & &
\\
$\calO_{ud\ell N1}^S(\star)$ & $(\overline{u_L}d_R)(\overline{\ell_L}N)$ & $n_f^2n_un_s$ &
$\calO_{ud\ell N2}^S(\star)$ & $(\overline{u_R}d_L)(\overline{\ell_L}N)$ & $n_f^2n_un_s$
\\
$\calO_{ud\ell N}^T(\star)$ & $(\overline{u_L}\sigma^{\mu\nu}d_R)(\overline{\ell_L}\sigma_{\mu\nu}N)$ & $n_f^2n_un_s $ &  & &
\\
$\calO_{\nu N\nu N}^S(\star\star)$ & $(\overline{\nu} N)(\overline{\nu}N)$ & $\frac{1}{2} n_fn_s(n_fn_s+1)$ &
& &
\\
\hline
\hline
 \multicolumn{6}{|c|}{$(\Delta L, \Delta B)=(2,~0)$}
\\
\hline
\multicolumn{3}{|c|}{\cellcolor{gray!35}$(\overline{L}L)(\overline{R}R)$}
&
\multicolumn{3}{|c|}{\cellcolor{gray!35}$(\overline{R}R)(\overline{R}R)$}
\\
\hline
$\calO_{\ell \nu N1}^V(\star)$ & $(\overline{\ell_L}\gamma^\mu \ell_L)(\overline{\nu^C}\gamma_\mu N)$ & $n_f^3n_s $ &
$\calO_{\ell \nu N2}^V(\star)$ & $(\overline{\ell_R}\gamma^\mu \ell_R)(\overline{\nu^C}\gamma_\mu N)$ & $n_f^3n_s  $
\\
$\calO_{d\nu N1}^V(\star)$ & $(\overline{d_L}\gamma^\mu d_L)(\overline{\nu^C}\gamma_\mu N)$ & $n_f^3n_s$ &
$\calO_{d\nu N2}^V(\star)$ & $(\overline{d_R}\gamma^\mu d_R)(\overline{\nu^C}\gamma_\mu N)$ & $n_f^3n_s$
\\
$\calO_{u\nu N1}^V(\star)$ & $(\overline{u_L}\gamma^\mu u_L)(\overline{\nu^C}\gamma_\mu N)$ & $n_fn_u^2n_s$ &
$\calO_{u\nu N2}^V(\star)$ & $(\overline{u_R}\gamma^\mu u_R)(\overline{\nu^C}\gamma_\mu N)$ & $n_fn_u^2n_s$
\\
$\calO_{du\ell N1}^V(\star)$ & $(\overline{d_L}\gamma^\mu u_L)(\overline{\ell_L^C}\gamma_\mu N)$ & $n_f^2n_un_s$ &
$\calO_{du\ell N2}^V(\star)$ & $(\overline{d_R}\gamma^\mu u_R)(\overline{\ell_L^C}\gamma_\mu N)$ & $n_f^2n_un_s$
\\
$\calO_{\nu\nu N}^V(\star)$ & $(\overline{\nu}\gamma^\mu \nu)(\overline{\nu^C}\gamma_\mu N)$ &
$\frac{1}{2}n_f^2(n_f+1)n_s$ &
$\calO_{N\nu N}^V(\star\star\star)$ & $(\overline{N}\gamma^\mu N)(\overline{\nu^C}\gamma_\mu N)$ &
$\frac{1}{2}n_fn_s^2(n_s+1)$
\\
\hline
\end{tabular}}
\caption{Dim-6 operator basis involving RH neutrinos $N$ in LNEFT. Here all operators are non-hermitian expect those with a (H) in the first sector. The number of $\star$ after each operator indicates the number of the RH neutrinos involved in the same operator.}
\label{tab:LNEFTA}
\end{table}

\begin{table}[!h]
\footnotesize
\center
\resizebox{\linewidth}{!}{
	\renewcommand{\arraystretch}{0.9}
\begin{tabular}{|l | l |  l|l | l | l|}
\hline
 Operator & Specific form & $\# (n_f,~n_u,~n_s)$ &  Operator & Specific form & $\# (n_f,~n_u,~n_s)$
\\
\hline
\multicolumn{3}{|c|}{\cellcolor{gray!35}$(\overline{L}R)(\overline{L}R)$}
&
\multicolumn{3}{|c|}{\cellcolor{gray!35}$(\overline{R}L)(\overline{L}R)$}
\\
\hline
$\calO_{\ell N1}^S(\star\star)$ & $(\overline{\ell_L}\ell_R)(\overline{N^C}N)$ & $\frac{1}{2}n_f^2n_s(n_s+1)$ &
$\calO_{\ell N2}^S(\star\star)$ & $(\overline{\ell_R}\ell_L)(\overline{N^C}N)$ & $\frac{1}{2}n_f^2n_s(n_s+1)$
\\
$\calO_{\ell N}^T(\star\star)$ & $(\overline{\ell_L}\sigma^{\mu\nu}\ell_R)(\overline{N^C}\sigma_{\mu\nu}N)$ & $\frac{1}{2}n_f^2n_s(n_s-1)$ &  & &
\\
\hline
$\calO_{dN1}^S(\star\star)$ & $(\overline{d_L}d_R)(\overline{N^C}N)$ & $\frac{1}{2}n_f^2n_s(n_s+1)$ &
$\calO_{dN2}^S(\star\star)$ & $(\overline{d_R}d_L)(\overline{N^C}N)$ & $\frac{1}{2}n_f^2n_s(n_s+1)$
\\
$\calO_{dN}^T(\star\star)$ & $(\overline{d_L}\sigma^{\mu\nu}d_R)(\overline{N^C}\sigma_{\mu\nu}N)$ & $\frac{1}{2}n_f^2n_s(n_s-1)$ &  & &
\\
$\calO_{uN1}^S(\star\star)$ & $(\overline{u_L}u_R)(\overline{N^C}N)$ & $\frac{1}{2}n_u^2n_s(n_s+1)$  &
$\calO_{uN2}^S(\star\star)$ & $(\overline{u_R}u_L)(\overline{N^C}N)$ & $\frac{1}{2}n_u^2n_s(n_s+1)$
\\
$\calO_{uN}^T(\star\star)$ & $(\overline{u_L}\sigma^{\mu\nu}u_R)(\overline{N^C}\sigma_{\mu\nu}N)$ & $\frac{1}{2}n_u^2n_s(n_s-1)$ &  & &
\\
$\calO_{du\ell N1}^S(\star)$& $(\overline{d_L}u_R)(\overline{\ell_R^C}N)$ & $n_f^2n_un_s$ &
$\calO_{du\ell N2}^S(\star)$ & $(\overline{d_R}u_L)(\overline{\ell_R^C}N)$ & $n_f^2n_un_s$
\\
\cline{4-6}
$\calO_{du\ell N}^T(\star)$& $(\overline{d_L}\sigma^{\mu\nu} u_R)(\overline{\ell_R^C}\sigma_{\mu\nu} N)$ & $n_f^2n_un_s$ &
\multicolumn{3}{|c|}{\cellcolor{gray!35}$(\overline{R}L)(\overline{R}L)$}
\\
\cline{4-6}
$\calO_{\nu NN}^S(\star\star\star)$& $(\overline{\nu}N)(\overline{N^C}N)$ & $\frac{1}{3}n_fn_s(n_s^2-1)$ &
$\calO_{N\nu \nu}^S(\star)$ & $(\overline{N}\nu )(\overline{\nu^C}\nu)$ & $\frac{1}{3}n_f(n_f^2-1)n_s$
\\
\hline
\hline
 \multicolumn{6}{|c|}{$(\Delta L, \Delta B)=(4,~0)$}
\\
\hline
\multicolumn{3}{|c|}{\cellcolor{gray!35}$(\overline{L}R)(\overline{L}R)$}
&
\multicolumn{3}{|c|}{\cellcolor{gray!35}$(\overline{R}L)(\overline{L}R)$}
\\
\hline
$\calO_{N}^S(\star\star\star\star)$ &$(\overline{N^C}N)(\overline{N^C}N)$  & $\frac{1}{12}n_s^2(n_s^2-1)$ &
$\calO_{\nu N}^S(\star\star)$ & $(\overline{\nu^C}\nu)(\overline{N^C}N)$ &  $\frac{1}{4}n_f(n_f+1)n_s(n_s+1)$
\\
\hline
 \multicolumn{6}{|c|}{$(\Delta L, \Delta B)=(1,~-1)$}
\\
\hline
\multicolumn{3}{|c|}{\cellcolor{gray!35}$(\overline{R}R)(\overline{R}R)$}
&
\multicolumn{3}{|c|}{\cellcolor{gray!35}$(\overline{L}R)(\overline{L}R)$}
\\
\hline
$\calO_{dduN1}^V(\star)$ & $(\overline{d_R}\gamma^\mu d_L^C)(\overline{u_R}\gamma_\mu N)$ & $n_f^2n_un_s$ &
$\calO_{uddN1}^S(\star)$ & $(\overline{u_L}d_L^C)(\overline{d_L} N)$ & $n_f^2n_un_s$
\\
\cline{1-3}
\multicolumn{3}{|c|}{\cellcolor{gray!35}$(\overline{R}L)(\overline{L}R)$} & & &
\\
\cline{1-3}
$\calO_{dduN1}^S(\star)$ & $(\overline{d_R}d_R^C)(\overline{u_L} N)$ &$\frac{1}{2}n_f(n_f-1)n_un_s$ & & &
\\
\hline
\hline
\multicolumn{6}{|c|}{$(\Delta L, \Delta B)=(1,~1)$}
\\
\hline
\multicolumn{3}{|c|}{\cellcolor{gray!35}$(\overline{L}L)(\overline{R}R)$}
&
\multicolumn{3}{|c|}{\cellcolor{gray!35}$(\overline{L}R)(\overline{L}R)$}
\\
\hline
$\calO_{dduN2}^V(\star)$ & $(\overline{d_R^C}\gamma^\mu d_L)(\overline{u_L^C}\gamma_\mu N)$ & $n_f^2n_un_s$ &
$\calO_{uddN2}^V(\star)$ & $(\overline{u_R^C}d_R)(\overline{d_R^C} N)$ & $n_f^2n_un_s$
\\
\cline{1-3}
\multicolumn{3}{|c|}{\cellcolor{gray!35}$(\overline{R}L)(\overline{L}R)$}  & & &
\\
\cline{1-3}
$\calO_{dduN2}^S(\star)$ & $(\overline{d_L^C} d_L)(\overline{u_R^C}N)$ & $\frac{1}{2}n_f(n_f-1)n_un_s$ & & &
\\
\hline
\hline
\multicolumn{6}{|c|}{Total $\#=2331|_{B=0}^{L=0} + 2304|_{B=0}^{L=2} +84|_{B=0}^{L=4}+252|_{B=-1}^{L=1} +252|_{B=1}^{L=2}=5223$,~~ $(n_f,~n_u,~n_s)=(3,~2,~3)$}
\\
\hline
\end{tabular}}
\caption{Continuation of Table~\ref{tab:LNEFTA}. }
\label{tab:LNEFTB}
\end{table}

\section{The SMNEFT operator basis at dim-6 and dim-7}
\label{sec:SMNEFTbasis}
Besides the SMEFT operators at dim-6~\cite{Grzadkowski:2010es} and dim-7~\cite{Lehman:2014jma,Liao:2016hru}, the SMNEFT also includes additional operators involving RH SM singlet fermions $N$. These operators with RH neutrino $N$ are classified in Ref.~\cite{Liao:2016qyd} and repeated in Table~\ref{tab:SMNEFT6} at dim-6 and Table~\ref{tab:SMNEFT7} at dim-7. For the dim-7 operators, by using the Fierz transformations here, we have rearranged some of the four-fermion operators given in Ref.~\cite{Liao:2016qyd} to have clear flavor symmetry and quark-lepton current structure. In addition, for the operators involving gauge field strength tensors, we accompany a corresponding gauge coupling constant for each involved field strength tensor.  Besides the operator basis involving RH neutrinos $N$ in Table~\ref{tab:SMNEFT6} and Table~\ref{tab:SMNEFT7}, in our matching calculation we also need the following relevant SMEFT dim-6 operators
\begin{align}\nonumber
&{\cal O}_{Hl}^{(3)}=(H^\dagger i\overleftrightarrow{D}^I_\mu H)(\overline{L}\tau^I\gamma^\mu L)\;,
&
&{\cal O}_{Hq}^{(3)}=(H^\dagger i\overleftrightarrow{D}^I_\mu H)(\overline{Q}\tau^I\gamma^\mu Q)\;,
\\\nonumber
&{\cal O}_{Hud}=i(\tilde H^\dagger D_\mu H)(\overline{u}\gamma^\mu d)\;,
&
&{\cal O}_{lq}^{(3)}=(\overline{L}\gamma_\mu \tau^I L)(\overline{Q}\gamma^\mu \tau^IQ)\;,
\\\nonumber
&{\cal O}_{ledq}=(\overline{L}e)(\overline{d}Q)\;,
&
&{\cal O}_{lequ}^{(1)}=(\overline{L}e)\epsilon(\overline{Q}u)\;,
\\
&{\cal O}_{lequ}^{(3)}=(\overline{L}\sigma_{\mu\nu}e)\epsilon (\overline{Q}\sigma^{\mu\nu}u)\;,
&&
\end{align}
and also dim-7 operators~\cite{Liao:2019tep}
\begin{align}\nonumber
&\mathcal{O}_{LeHD}=\epsilon_{ij}\epsilon_{mn}(\overline{L^{C,i}}\gamma_\mu e)H^j(H^miD^\mu H^n)\;,
&&
\\\nonumber
&\mathcal{O}_{\overline{d}QLLH1}=\epsilon_{ij}\epsilon_{mn}(\overline{d}Q^i)(\overline{L^{C,j}}L^m)H^n\;,
&&\mathcal{O}_{\overline{d}uLeH}=\epsilon_{ij}(\overline{d}\gamma_\mu u)(\overline{L^{C,i}}\gamma^\mu e)H^j\;,
\\
&\mathcal{O}_{\overline{d}QLLH2}=\epsilon_{ij}\epsilon_{mn}(\overline{d}\sigma_{\mu\nu}Q^i)
(\overline{L^{C,j}}\sigma^{\mu\nu} L^m)H^n\;,
&&\mathcal{O}_{\overline{Q}uLLH}=\epsilon_{ij}(\overline{Q}u)(\overline{L^{C}}L^i)H^j\;.
\end{align}

\begin{table}[!h]
\center
\parbox{1\linewidth}{
\scriptsize
\centering
\setlength{\tabcolsep}{9.5pt}
\renewcommand{\arraystretch}{1.1}
\begin{tabular}{|c|c|c|c|c|c|}
\hline
\multicolumn{2}{|c|}{$\psi^2H^3(+\mbox{h.c.})$}
&\multicolumn{2}{|c|}{$(\overline{L}R)(\overline{L}R)(+\mbox{h.c.})$}
&\multicolumn{2}{|c|}{$(\overline{L}L)(\overline{R}R)$}
\\\hline
$\calO_{LNH}$ & $(\overline{L}N)\tilde{H}(H^\dagger H)$
& $\calO_{LNLe}$ & $(\overline{L}N)\epsilon(\overline{L}e)$
&$\calO_{LN}$ & $(\overline{L}\gamma^\mu L)(\overline{N}\gamma_\mu N)$
\\\cline{1-4}
\multicolumn{2}{|c|}{$\psi^2H^2D (+\mbox{h.c.})$}
&$\calO_{LNQd}$ & $(\overline{L}N)\epsilon(\overline{Q}d)$
&$\calO_{QN}$ & $(\overline{Q}\gamma^\mu Q)(\overline{N}\gamma_\mu N)$
\\\cline{1-2}\cline{5-6}
$\calO_{HN}(\text{H})$ &  $(\overline{N}\gamma^\mu N)(H^\dagger i \overleftrightarrow{D_\mu} H)$
& $\calO_{LdQN}$ & $(\overline{L}d)\epsilon(\overline{Q}N)$
&\multicolumn{2}{|c|}{\cellcolor{gray!35}$(\Delta L, \Delta B)=(4,~0)$}
\\\cline{3-6}
$\calO_{HNe}$ & $(\overline{N}\gamma^\mu e)({\tilde{H}}^\dagger i D_\mu H)$
& \multicolumn{2}{|c|}{$(\overline{R}R)(\overline{R}R)$}
&$\cellcolor{gray!35}\calO_{NNNN}$ &$\cellcolor{gray!35}(\overline{N^C}N)(\overline{N^C}N)$
\\\hline
\multicolumn{2}{|c|}{$\psi^2HX(+\mbox{h.c.})$}
&$\calO_{NN}$ & $(\overline{N}\gamma^\mu N)(\overline{N}\gamma_\mu N)$
&\multicolumn{2}{|c|}{$\cellcolor{gray!35}(\Delta L, \Delta B)=(1,~1)$}
\\\cline{1-2}\cline{5-6}
$\calO_{NB}$ & $g_1(\overline{L}\sigma_{\mu\nu}N)\tilde{H}B^{\mu\nu}$
&$\calO_{eN}$ & $(\overline{e}\gamma^\mu e)(\overline{N}\gamma_\mu N)$
&$\cellcolor{gray!35}\calO_{QQdN}$ &$\cellcolor{gray!35}\epsilon_{ij}\epsilon_{\alpha\beta\sigma}(\overline{Q^{i,C}_{\alpha}}Q^j_{\beta})(\overline{d_{\sigma}^C}N)$
\\
$\calO_{NW}$ &$g_2(\overline{L}\sigma_{\mu\nu}N)\tau^I\tilde{H}W^{I\mu\nu}$
&$\calO_{uN}$& $(\overline{u}\gamma^\mu u)(\overline{N}\gamma_\mu N)$
&$\cellcolor{gray!35}\calO_{uddN}$ &$\cellcolor{gray!35}\epsilon_{\alpha\beta\sigma}(\overline{u_{\alpha}^C}d_{\beta})(\overline{d_{\sigma}^C}N)$
\\\cline{1-2}\cline{5-6}
\multicolumn{2}{|c|}{$(\overline{L}R)(\overline{R}L)(+\mbox{h.c.})$}
&$\calO_{dN}$& $(\overline{d}\gamma^\mu d)(\overline{N}\gamma_\mu N)$
& &
\\\cline{1-2}
$\calO_{QuNL}$& $(\overline{Q}u)(\overline{N}L)$
&$\calO_{duNe}(+\mbox{h.c.})$& $ (\overline{d}\gamma^\mu u)(\overline{N}\gamma_\mu e)$
& &
\\
\hline
\end{tabular}
\caption{The basis of dim-6 operators involving RH neutrino $N$ in SMNEFT~\cite{Liao:2016hru}, where $\alpha,~\beta,~\sigma$ and $i,~j$ are $SU(3)_C$ and $SU(2)_L$ indices, respectively.}
\label{tab:SMNEFT6}
}
\vspace{0.5cm}
\\
\parbox{1\linewidth}{
\scriptsize
\centering
\setlength{\tabcolsep}{3pt}
\renewcommand{\arraystretch}{1.1}
\begin{tabular}{|c|c|c|c|c|c|}
\hline
 \multicolumn{2}{|c|}{$N\psi H^3D$} &  \multicolumn{2}{|c|}{$N\psi^3D$}   &   \multicolumn{2}{|c|}{$N^2\psi^2H$}
\\\hline
$\mathcal{O}_{NL1}$ & $\epsilon_{ij}(\overline{N^C}\gamma_\mu L^i) (iD^\mu H^j)(H^\dagger H)$
&$\mathcal{O}_{eNLLD}$ & $\epsilon_{ij}(\overline{e}\gamma_\mu N)(\overline{L^{i,C}}i\overleftrightarrow{D}^\mu L^j)$
& $\mathcal{O}_{LNeH}$ & $(\overline{L}N)(\overline{N^C}e)H$
\\
$\mathcal{O}_{NL2}$&$\epsilon_{ij}(\overline{N^C}\gamma_\mu L^i)  H^j(H^\dagger i\overleftrightarrow{D^\mu} H)$
&  $\mathcal{O}_{duNeD}$& $(\overline{d}\gamma_\mu u)(\overline{N^C}i\overleftrightarrow{D}^\mu e)$
& $\mathcal{O}_{eLNH}$ & $H^\dagger(\overline{e}L)(\overline{N^C}N)$
\\\cline{1-2}
 \multicolumn{2}{|c|}{$N\psi H^2D^2$}   &  $\mathcal{O}_{QuNLD}$ & $(\overline{Q}i\overleftrightarrow{D}_\mu u)(\overline{N^C}\gamma^\mu L)$ &
$\mathcal{O}_{QNdH}$ & $(\overline{Q}N)(\overline{N^C}d)H$
\\\cline{1-2}
$\mathcal{O}_{NeD}$ &  $\epsilon_{ij}(\overline{N^C}\overleftrightarrow{D}^\mu e)(H^iD^\mu H^j)$
&  $\mathcal{O}_{dQNLD}$  & $\epsilon_{ij}(\overline{d}i\overleftrightarrow{D}_\mu Q^i )(\overline{N^C}\gamma^\mu L^j) $
&  $\mathcal{O}_{dQNH}$ & $H^\dagger(\overline{d}Q)(\overline{N^C}N)$
\\\cline{1-4}
\multicolumn{2}{|c|}{$N\psi H^2X$}
& \multicolumn{2}{|c|}{$N^2\psi^2D$}
& $\mathcal{O}_{QNuH}$ & $(\overline{Q}N)(\overline{N^C}u)\tilde{H}$
\\\cline{1-4}
$\mathcal{O}_{NeW}$ & $g_2(\epsilon\tau^I)_{ij}(\overline{N^C}\sigma^{\mu\nu}e)(H^iH^j)W^I_{\mu\nu}$
& $\mathcal{O}_{LND}$ & $(\overline{L}\gamma_\mu L)(\overline{N^C}i\overleftrightarrow{\partial}^\mu N)$
& $\mathcal{O}_{uQNH}$ & $\tilde{H}^\dagger(\overline{u}Q)(\overline{N^C}N)$
\\\cline{1-2}\cline{5-6}
\multicolumn{2}{|c|}{$N\psi HDX$}
&  $\mathcal{O}_{QND}$ & $(\overline{Q}\gamma_\mu Q)(\overline{N^C}i\overleftrightarrow{\partial}^\mu N)$
&  \multicolumn{2}{|c|}{$N^3\psi H$}
\\\cline{1-2}\cline{5-6}
$\mathcal{O}_{NLB1}$ & $g_1\epsilon_{ij}(\overline{N^C}\gamma^\mu L^i)(D^\nu H^j)B_{\mu\nu}$
& $\mathcal{O}_{eND}$ & $(\overline{e}\gamma_\mu e)(\overline{N^C}i\overleftrightarrow{\partial}^\mu N)$
& $\mathcal{O}_{LNNH}$  & $(\overline{L}N)(\overline{N^C}N)\tilde{H}$
\\
$\mathcal{O}_{NLB2}$ & $g_1\epsilon_{ij}(\overline{N^C}\gamma^\mu L^i)(D^\nu H^j)\tilde{B}_{\mu\nu}$
& $\mathcal{O}_{uND}$ & $(\overline{u}\gamma_\mu u)(\overline{N^C}i\overleftrightarrow{\partial}^\mu N)$
& $\mathcal{O}_{NLNH}$ & $\tilde{H}^\dagger(\overline{N}L)(\overline{N^C}N)$
\\\cline{5-6}
$\mathcal{O}_{NLW1}$ & $g_2(\epsilon \tau^I)_{ij}(\overline{N^C}\gamma^\mu L^i)(D^\nu H^j)W^I_{\mu\nu}$
& $\mathcal{O}_{dND}$ & $(\overline{d}\gamma_\mu d)(\overline{N^C}i\overleftrightarrow{\partial}^\mu N)$
&  \multicolumn{2}{|c|}{\cellcolor{gray!35}$\slashed{B}:~N\psi^3D~\&~N\psi^3H$}
\\\cline{3-6}
$\mathcal{O}_{NLW2}$ & $g_2(\epsilon \tau^I)_{ij}(\overline{N^C}\gamma^\mu L^i)(D^\nu H^j)\tilde{W}^I_{\mu\nu}$
&\multicolumn{2}{|c|}{$N^4D$}
& $\cellcolor{gray!35} \mathcal{O}_{uNdD}$ &\cellcolor{gray!35}$\epsilon_{\alpha\beta\sigma}(\overline{u}_{\alpha}\gamma_\mu N)(\overline{d}_{\beta}i\overleftrightarrow{D}^\mu d^C_{\sigma})$
\\\cline{1-4}
\multicolumn{2}{|c|}{$N^2H^4$}
& $\mathcal{O}_{NND}$ & $(\overline{N}\gamma_\mu N)(\overline{N^C}i\overleftrightarrow{\partial}^\mu N) $
& $\cellcolor{gray!35} \mathcal{O}_{dNQD}$ &\cellcolor{gray!35}$\epsilon_{ij}\epsilon_{\alpha\beta\sigma}(\overline{d}_{\alpha}\gamma_\mu N) (\overline{Q}_{i\beta}i\overleftrightarrow{D}^\mu Q_{j\sigma}^C)$
\\\cline{1-4}
$\mathcal{O}_{NH}$ & $(\overline{N^C}N)(H^\dagger H)^2$
&\multicolumn{2}{|c|}{$N\psi^3 H$}
&\cellcolor{gray!35}$\mathcal{O}_{QNdH}$ &\cellcolor{gray!35}$\epsilon_{ij}\epsilon_{\alpha\beta\sigma}(\overline{Q}_{i\alpha}N)(\overline{d}_\beta d^C_\sigma)\tilde{H}^j$
\\\cline{1-4}
\multicolumn{2}{|c|}{$N^2H^2D^2$}
& $\mathcal{O}_{LNLH}$ & $\epsilon_{ij}(\overline{L}\gamma_\mu L)(\overline{N^C}\gamma^\mu L^i)H^j$
 &\cellcolor{gray!35}$\mathcal{O}_{QNQH}$ &\cellcolor{gray!35}$\epsilon_{ij}\epsilon_{\alpha\beta\sigma}(\overline{Q}_{i\alpha}N)(\overline{Q}_{j\beta }Q^C_\sigma)H$
\\\cline{1-2}
$\mathcal{O}_{NHD1}$ & $(\overline{N^C}\overleftrightarrow{\partial}_\mu N)(H^\dagger \overleftrightarrow{D^\mu} H)$
& $\mathcal{O}_{QNLH1}$ & $\epsilon_{ij}(\overline{Q}\gamma_\mu Q)(\overline{N^C}\gamma^\mu L^i)H^j$
&\cellcolor{gray!35}$\mathcal{O}_{QNudH}$ &\cellcolor{gray!35}$\epsilon_{\alpha\beta\sigma}(\overline{Q}_{\alpha}N)(\overline{u}_\beta d^C_\sigma)H$
\\\cline{5-6}
    $\mathcal{O}_{NHD2}$        & $(\overline{N^C} N)(D_\mu H)^\dagger D^\mu H$
& $\mathcal{O}_{QNLH2}$ & $\epsilon_{ij}(\overline{Q}\gamma_\mu Q^i)(\overline{N^C}\gamma^\mu L^j)H$
&\multicolumn{2}{|c|}{$N^2X^2$}
\\\cline{1-2}\cline{5-6}
\multicolumn{2}{|c|}{$N^2H^2X$}
&$\mathcal{O}_{eNLH}$ & $\epsilon_{ij}(\overline{e}\gamma_\mu e )(\overline{N^C} \gamma^\mu L^i)H^j$
& $\mathcal{O}_{NB1}$ & $\alpha_1(\overline{N^C}N)B_{\mu\nu}B^{\mu\nu}$
\\\cline{1-2}
$\mathcal{O}_{NHB}$ & $g_1(\overline{N^C}\sigma_{\mu\nu}N)(H^\dagger H)B^{\mu\nu}$
& $\mathcal{O}_{dNLH}$   & $\epsilon_{ij}(\overline{d}\gamma_\mu d)(\overline{N^C}\gamma^\mu L^i)H^j$
& $\mathcal{O}_{NB2}$        & $\alpha_1(\overline{N^C}N)B_{\mu\nu}\tilde{B}^{\mu\nu}$
\\
$\mathcal{O}_{NHW}$       & $g_2(\overline{N^C}\sigma_{\mu\nu}N)(H^\dagger\tau^I H)W^{I\mu\nu}$
& $\mathcal{O}_{uNLH}$ & $\epsilon_{ij}(\overline{u}\gamma_\mu u)(\overline{N^C} \gamma^\mu L^i)H^j$
& $\mathcal{O}_{NW1}$    & $\alpha_2(\overline{N^C}N)W^I_{\mu\nu}W^{I\mu\nu}$
\\
&
&  $\mathcal{O}_{duNLH}$ & $\epsilon_{ij}(\overline{d}\gamma_\mu u)(\overline{N^C} \gamma^\mu L^i)\tilde{H}^j$
&  $\mathcal{O}_{NW2}$ & $\alpha_2(\overline{N^C}N)W^I_{\mu\nu}\tilde{W}^{I\mu\nu}$
\\
&
& $\mathcal{O}_{dQNeH}$ & $\epsilon_{ij}(\overline{d}Q^i)(\overline{N^C}e)H^j$
& $\mathcal{O}_{NG1}$      & $\alpha_3(\overline{N^C}N)G^A_{\mu\nu}G^{A\mu\nu}$
\\
&
& $\mathcal{O}_{QuNeH1}$ & $(\overline{Q}u)(\overline{N^C}e)H$
& $\mathcal{O}_{NG2}$      & $\alpha_3(\overline{N^C}N)G^A_{\mu\nu}\tilde{G}^{A\mu\nu}$
\\
&
& $\mathcal{O}_{QuNeH2}$ & $(\overline{Q}\sigma_{\mu\nu}u)(\overline{N^C}\sigma^{\mu\nu}e)H$
& &
\\\hline
\end{tabular}
\caption{The basis of dim-7 operators involving RH neutrino $N$ in SMNEFT, where all of the operators are non-hermitian with the net global quantum number $|\Delta L-\Delta B|=2$. Here $g_{1,2,3}$ are the gauge coupling constants for the gauge groups $U(1)_Y, SU(2)_L,SU(3)_C$, respectively, and $\alpha_i=g_i^2/(4\pi)$.
}
\label{tab:SMNEFT7}
}
\end{table}

\section{The decay matrix elements of meson and tau lepton}
\label{sec:mesonmatrix}
For $P^+(u_p\bar{d}_r)\to \ell^+_\alpha \nu/N$, we find the following $\Delta L=0$ hadron-level amplitudes
\begin{eqnarray}
\mathcal{M}(P^+\to \ell^+_\alpha \nu_\beta)&=&-{i\over 2}f_P \Big(C_{ud\ell\nu 1}^{V\ast}- C_{ud\ell\nu 2}^{V\ast}\Big)\overline{u_\nu}\cancel{p} P_L v_{\ell^+}\nonumber \\
&-&{im_P^2 f_P\over 2(m_{u_p}+m_{d_r})} \Big( C_{ud\ell\nu 1}^{S\ast}-C_{ud\ell\nu 2}^{S\ast}\Big)\overline{u_\nu} P_R v_{\ell^+}\;,\\
\mathcal{M}(P^+\to \ell^+_\alpha N_\beta)&=&-{i\over 2}f_P \Big(C_{ud\ell N 1}^{V\ast}- C_{ud\ell N 2}^{V\ast}\Big)\overline{u_N}\cancel{p} P_R v_{\ell^+}\nonumber \\
&+&{im_P^2 f_P\over 2(m_{u_p}+m_{d_r})} \Big(C_{ud\ell N 1}^{S\ast}-C_{ud\ell N 2}^{S\ast}\Big)\overline{u_N} P_L v_{\ell^+}\;,
\end{eqnarray}
where $p^\mu$ denotes the momentum of the meson. The $\Delta L=-2$ amplitudes for $P^+(u_p\bar{d}_r)\to \ell^+_\alpha \bar\nu/\bar N$ are
\begin{eqnarray}
\mathcal{M}(P^+\to \ell^+_\alpha \bar{\nu}_\beta)&=&-{i\over 2}f_P\Big(C_{du\ell\nu 1}^{V}-C_{du\ell\nu 2}^{V}\Big)\overline{v_{\ell^+}^C}\cancel{p} P_L v_{\bar{\nu}}\nonumber \\
&+&{im_P^2 f_P\over 2(m_{u_p}+m_{d_r})}\Big( C_{du\ell\nu 1}^{S} -C_{du\ell\nu 2}^{S}\Big)\overline{v_{\ell^+}^C} P_L v_{\bar{\nu}}\;,\\
\mathcal{M}(P^+\to \ell^+_\alpha \bar{N}_\beta)&=&-{i\over 2}f_P\Big(C_{du\ell N1}^{V} - C_{du\ell N2}^{V}\Big)\overline{v_{\ell^+}^C} \cancel{p} P_R v_{\bar{N}}\nonumber \\
&-&{im_P^2 f_P\over 2(m_{u_p}+m_{d_r})}\Big(C_{du\ell N1}^{S}- C_{du\ell N2}^{S}\Big)\overline{v_{\ell^+}^C} P_R v_{\bar{N}} \;.
\end{eqnarray}
For $V^+(u_p\bar{d}_r)\to \ell^+ \nu/N$, the LNC hadron-level matrix elements with $\Delta L=0$ are
\begin{eqnarray}
\mathcal{M}(V^+\to \ell^+_\alpha \nu_\beta)&=&{1\over 2}\Big(C_{ud\ell\nu 1}^{V\ast}+C_{ud\ell\nu 2}^{V\ast}\Big)f_V m_V \epsilon_V^\mu\overline{u_\nu}\gamma_\mu P_L v_{\ell^+}\nonumber \\
&+&i C_{ud\ell\nu}^{T\ast}f_V^{T}(\epsilon_V^\mu p_V^\nu-\epsilon_V^\nu p_V^\mu)\overline{u_\nu} \sigma_{\mu\nu} P_R v_{\ell^+}\;,\\
\mathcal{M}(V^+\to \ell^+_\alpha N_\beta)&=&{1\over 2}\Big( C_{ud\ell N1}^{V\ast}+C_{ud\ell N2}^{V\ast}\Big)f_V m_V \epsilon_V^\mu\overline{u_N}\gamma^\mu P_R v_{\ell^+}\nonumber \\
&+&i C_{ud\ell N}^{T\ast}f_V^{T}(\epsilon_V^\mu p_V^\nu-\epsilon_V^\nu p_V^\mu)\overline{u_N} \sigma_{\mu\nu} P_L v_{\ell^+} \;.
\end{eqnarray}
The LNV matrix elements with $\Delta L=-2$ for $V^+(u_p\bar{d}_r)\to \ell^+ \bar{\nu}/\bar N$ are
\begin{eqnarray}
\mathcal{M}(V^+\to \ell^+_\alpha \bar{\nu}_\beta)&=&{1\over 2}\Big( C_{du\ell\nu 1}^{V}+C_{du\ell\nu 2}^{V}\Big)f_V m_V \epsilon_V^\mu\overline{v_{\ell^+}^C}\gamma_\mu P_L v_{\bar{\nu}}\nonumber \\
&+&i C_{du\ell\nu}^{T}f_V^{T}(\epsilon_V^\mu p_V^\nu-\epsilon_V^\nu p_V^\mu)\overline{v_{\ell^+}^C} \sigma_{\mu\nu} P_L v_{\bar{\nu}} \;,\\
\mathcal{M}(V^+\to \ell^+_\alpha \bar{N}_\beta)&=&{1\over 2}\Big( C_{du\ell N1}^{V}+C_{du\ell N2}^{V}\Big)f_V m_V \epsilon_V^\mu\overline{v_{\ell^+}^C}\gamma_\mu P_R v_{\bar{N}}\nonumber \\
&+&i C_{du\ell N}^{T}f_V^{T}(\epsilon_V^\mu p_V^\nu-\epsilon_V^\nu p_V^\mu)\overline{v_{\ell^+}^C} \sigma_{\mu\nu} P_R v_{\bar{N}} \;.
\end{eqnarray}
For $\tau^-\to P^-(\bar{u}_p d_r) \nu/\bar N$, the LNC hadron-level matrix elements with $\Delta L=0$ are
\begin{eqnarray}
\mathcal{M}(\tau^-\to P^- \nu_\beta)&=&{i\over 2}f_P \Big(C_{ud\ell\nu 1}^{V\ast}-C_{ud\ell\nu 2}^{V\ast}\Big)\overline{u_\nu}\cancel{p} P_L u_{\tau^-}\nonumber \\
&-&{im_P^2 f_P \over 2(m_{u_p}+m_{d_r})}\Big( C_{ud\ell\nu 1}^{S\ast}-C_{ud\ell\nu 2}^{S\ast}\Big)\overline{u_\nu} P_R u_{\tau^-}\;,\\
\mathcal{M}(\tau^-\to P^-N_\beta)&=&{i\over 2}f_P \Big(C_{ud\ell N1}^{V\ast} -C_{ud\ell N2}^{V\ast}\Big)\overline{u_N} \cancel{p} P_R u_{\tau^-}\nonumber \\
&+&{im_P^2 f_P\over 2(m_{u_p}+m_{d_r})}\Big(C_{ud\ell N1}^{S\ast}-C_{ud\ell N2}^{S\ast}\Big)\overline{u_N} P_L u_{\tau^-}\;.
\end{eqnarray}
For $\tau^-\to P^-(\bar{u}_p d_r) \bar{\nu}/\bar N$, the LNV hadron-level matrix elements with $\Delta L=-2$ are
\begin{eqnarray}
\mathcal{M}(\tau^-\to P^-\bar{\nu}_\beta)&=&{i\over 2}f_P \Big(C_{du\ell\nu 1}^{V} -C_{du\ell\nu 2}^{V}\Big)\overline{u_{\tau^-}^C} \cancel{p} P_L v_{\bar{\nu}}\nonumber \\
&+&{im_P^2 f_P\over 2(m_{u_p}+m_{d_r})}\Big(C_{du\ell\nu 1}^{S} -C_{du\ell\nu 2}^{S}\Big)\overline{u_{\tau^-}^C} P_L v_{\bar{\nu}}\;,\\
\mathcal{M}(\tau^-\to P^-\bar{N}_\beta)&=&{i\over 2}f_P\Big(C_{du\ell N1}^{V}-C_{du\ell N2}^{V}\Big)\overline{u_{\tau^-}^C} \cancel{p} P_R v_{\bar{N}}\nonumber \\
&-&{im_P^2 f_P\over 2(m_{u_p}+m_{d_r})}\Big(C_{du\ell N1}^{S}- C_{du\ell N2}^{S}\Big)\overline{u_{\tau^-}^C} P_R v_{\bar{N}}\;.
\end{eqnarray}
For $\tau^-\to V^-(\bar{u}_p d_r) \nu/N$, the LNC hadron-level matrix elements with $\Delta L=0$ are
\begin{eqnarray}
\mathcal{M}(\tau^-\to V^-\nu_\beta)&=&{1\over 2}\Big( C_{ud\ell\nu 1}^{V\ast}+C_{ud\ell\nu 2}^{V\ast}\Big)f_V m_V \epsilon_V^{\mu\ast}\overline{u_\nu}\gamma_\mu P_L u_{\tau^-}\nonumber \\
&-&i C_{ud\ell\nu}^{T\ast}f_V^{T}(\epsilon_V^{\mu\ast} p_V^\nu-\epsilon_V^{\nu\ast} p_V^\mu)\overline{u_\nu} \sigma_{\mu\nu} P_R u_{\tau^-}\;,\\
\mathcal{M}(\tau^-\to V^-N_\beta)&=&{1\over 2}\Big( C_{ud\ell N1}^{V\ast}+C_{ud\ell N2}^{V\ast}\Big)f_V m_V \epsilon_V^{\mu\ast}\overline{u_N}\gamma_\mu P_R u_{\tau^-}\nonumber \\
&-&i C_{ud\ell N}^{T\ast}f_V^{T}(\epsilon_V^{\mu\ast} p_V^\nu-\epsilon_V^{\nu\ast} p_V^\mu)\overline{u_N} \sigma_{\mu\nu} P_L u_{\tau^-} \;.
\end{eqnarray}
For $\tau^-\to V^-(\bar{u}_p d_r) \bar{\nu}/\bar N$, the LNV hadron-level matrix elements with $\Delta L=-2$ are
\begin{eqnarray}
\mathcal{M}(\tau^-\to V^-\bar{\nu}_\beta)&=&{1\over 2}\Big( C_{du\ell\nu 1}^{V}+C_{du\ell\nu 2}^{V}\Big)f_V m_V \epsilon_V^{\mu\ast}\overline{u_{\tau^-}^C}\gamma_\mu P_L v_{\bar{\nu}}\nonumber \\
&-&i C_{du\ell\nu}^{T}f_V^{T}(\epsilon_V^{\mu\ast} p_V^\nu-\epsilon_V^{\nu\ast} p_V^\mu)\overline{u_{\tau^-}^C} \sigma_{\mu\nu} P_L v_{\bar{\nu}}\;,\\
\mathcal{M}(\tau^-\to V^-\bar{N}_\beta)&=&{1\over 2}\Big( C_{du\ell N1}^{V}+C_{du\ell N2}^{V}\Big)f_V m_V \epsilon_V^{\mu\ast}\overline{u_{\tau^-}^C}\gamma^\mu P_R v_{\bar{N}}\nonumber \\
&-&i C_{du\ell N}^{T}f_V^{T}(\epsilon_V^{\mu\ast} p_V^\nu-\epsilon_V^{\nu\ast} p_V^\mu)\overline{u_{\tau^-}^C} \sigma_{\mu\nu} P_R v_{\bar{N}} \;.
\end{eqnarray}

\section{$\beta$ decay}
\label{sec:betadecay}

Recently the authors of Ref.~\cite{Gonzalez-Alonso:2018omy} studied $\beta$ decay. We use their result and translate it to the full LNEFT operator basis. In this section, we collect the relevant matrix elements and report the translation to the operator basis used in Ref.~\cite{Gonzalez-Alonso:2018omy}. The neutron-to-proton transition can be parameterized as~\cite{Weinberg:1958ut,Bhattacharya:2011qm,Gonzalez-Alonso:2018omy}
\begin{eqnarray}\nonumber
\langle p(p_p)|\bar{u}\gamma_\mu d|n(p_n)\rangle &=& \bar{u}_p(p_p)g_V \gamma_\mu u_n(p_n) + \mathcal{O}(q/M_N)\;,\\\nonumber
\langle p(p_p)|\bar{u}\gamma_\mu \gamma_5 d|n(p_n)\rangle &=& \bar{u}_p(p_p)g_A \gamma_\mu \gamma_5 u_n(p_n) + \mathcal{O}(q/M_N)\;,\\\nonumber
\langle p(p_p)|\bar{u}d|n(p_n)\rangle &=& \bar{u}_p(p_p)g_S u_n(p_n) + \mathcal{O}(q^2/M_N^2)\;,\\\nonumber
\langle p(p_p)|\bar{u}\gamma_5 d|n(p_n)\rangle &=& \bar{u}_p(p_p)g_P\gamma_5 u_n(p_n) + \mathcal{O}(q^2/M_N^2)\;,\\
\langle p(p_p)|\bar{u}\sigma_{\mu\nu}d|n(p_n)\rangle &=& \bar{u}_p(p_p)g_T\sigma_{\mu\nu} u_n(p_n) + \mathcal{O}(q/M_N)
\end{eqnarray}
where the 4-momentum $q=p_n-p_p$ denotes the difference between the 4-momenta of the neutron and proton and $M_N=M_n=M_p$ denotes the nucleon mass. The numerical values of the nuclear form factors are~\cite{Gonzalez-Alonso:2018omy}
\begin{align}
g_V=&1\;, &
g_A=&1.278(33)\;, &
g_T=&0.987(55)\;, &
g_S=&1.02(11)\;, &
g_P=&349(9)\;.
\end{align}
The nucleon-level $\Delta L=0$ amplitudes are
\begin{eqnarray}
\mathcal{M}(n\to p \ell^-_\alpha \bar{\nu}_\beta )&=&{1\over 2}\Big(C_{ud\ell\nu 1}^{V}+C_{ud\ell\nu 2}^{V}\Big)g_V \bar{u}_p \gamma_\mu u_n \overline{u_{\ell^-}}\gamma^\mu P_L v_{\bar{\nu}}\nonumber \\
&-&{1\over 2}\Big(C_{ud\ell\nu 1}^{V}-C_{ud\ell\nu 2}^{V}\Big)g_A \bar{u}_p \gamma_\mu\gamma_5 u_n \overline{u_{\ell^-}}\gamma^\mu P_L v_{\bar{\nu}}\nonumber \\
&+&{1\over 2}\Big(C_{ud\ell\nu 1}^{S}+C_{ud\ell\nu 2}^{S}\Big)g_S \bar{u}_p u_n \overline{u_{\ell^-}} P_L v_{\bar{\nu}}\nonumber \\
&-&{1\over 2}\Big(C_{ud\ell\nu 1}^{S}-C_{ud\ell\nu 2}^{S}\Big)g_P \bar{u}_p \gamma_5 u_n \overline{u_{\ell^-}} P_L v_{\bar{\nu}}\nonumber \\
&+&C_{ud\ell\nu}^{T} g_T\bar{u}_p \sigma_{\mu\nu} u_n~\overline{u_{\ell^-}} \sigma^{\mu\nu} P_L v_{\bar{\nu}}\;,\\
\mathcal{M}(n\to p \ell^-_\alpha \bar{N}_\beta )&=&{1\over 2}\Big(C_{ud\ell N 1}^{V}+C_{ud\ell N 2}^{V}\Big)g_V \bar{u}_p \gamma_\mu u_n \overline{u_{\ell^-}}\gamma^\mu P_R v_{\bar{N}}\nonumber \\
&-&{1\over 2}\Big(C_{ud\ell N 1}^{V}-C_{ud\ell N 2}^{V}\Big)g_A \bar{u}_p \gamma_\mu\gamma_5 u_n \overline{u_{\ell^-}}\gamma^\mu P_R v_{\bar{N}}\nonumber \\
&+&{1\over 2}\Big(C_{ud\ell N 1}^{S}+C_{ud\ell N 2}^{S}\Big)g_S \bar{u}_p u_n \overline{u_{\ell^-}} P_R v_{\bar{N}}\nonumber \\
&+&{1\over 2}\Big(C_{ud\ell N 1}^{S}-C_{ud\ell N 2}^{S}\Big)g_P \bar{u}_p \gamma_5 u_n \overline{u_{\ell^-}} P_R v_{\bar{N}}\nonumber \\
&+&C_{ud\ell N}^{T} g_T\bar{u}_p \sigma_{\mu\nu} u_n~\overline{u_{\ell^-}} \sigma^{\mu\nu} P_R v_{\bar{N}}\;.
\end{eqnarray}
The $\Delta L=2$ amplitudes are
\begin{eqnarray}
\mathcal{M}(n\to p \ell^-_\alpha \nu_\beta)&=&{1\over 2}\Big(C_{du\ell\nu 1}^{V\ast} +C_{du\ell\nu 2}^{V\ast}\Big)g_V \bar{u}_p \gamma_\mu u_n \overline{u_{\nu}}\gamma^\mu P_L u_{\ell^-}^C\nonumber \\
&-&{1\over 2}\Big(C_{du\ell\nu 1}^{V\ast}-C_{du\ell\nu 2}^{V\ast}\Big)g_A \bar{u}_p \gamma_\mu \gamma_5 u_n \overline{u_{\nu}}\gamma^\mu P_L u_{\ell^-}^C\nonumber \\
&+&{1\over 2}\Big(C_{du\ell\nu 1}^{S\ast} +C_{du\ell\nu 2}^{S\ast}\Big)g_S \bar{u}_p u_n\overline{u_{\nu}} P_R u_{\ell^-}^C\nonumber \\
&+&{1\over 2}\Big(C_{du\ell\nu 1}^{S\ast} -C_{du\ell\nu 2}^{S\ast}\Big)g_P \bar{u}_p \gamma_5 u_n\overline{u_{\nu}} P_R u_{\ell^-}^C\nonumber \\
&+&C_{du\ell\nu}^{T\ast} g_T\bar{u}_p \sigma_{\mu\nu} u_n~\overline{u_{\nu}} \sigma^{\mu\nu} P_R u_{\ell^-}^C \;,\\
\mathcal{M}(n\to p \ell^-_\alpha N_\beta)&=&{1\over 2}\Big(C_{du\ell N1}^{V\ast} +C_{du\ell N2}^{V\ast}\Big) g_V \bar{u}_p \gamma_\mu u_n \overline{u_{N}}\gamma^\mu P_R u_{\ell^-}^C\nonumber \\
&-&{1\over 2}\Big(C_{du\ell N1}^{V\ast} +C_{du\ell N2}^{V\ast}\Big) g_A \bar{u}_p \gamma_\mu \gamma_5 u_n \overline{u_{N}}\gamma^\mu P_R u_{\ell^-}^C\nonumber \\
&+&{1\over 2}\Big(C_{du\ell N1}^{S\ast} +C_{du\ell N2}^{S\ast}\Big)g_S \bar{u}_p u_n \overline{u_{N}} P_L u_{\ell^-}^C\nonumber \\
&-&{1\over 2}\Big(C_{du\ell N1}^{S\ast}-C_{du\ell N2}^{S\ast}\Big)g_P \bar{u}_p \gamma_5 u_n \overline{u_{N}} P_L u_{\ell^-}^C\nonumber \\
&+&C_{du\ell N}^{T\ast}g_T\bar{u}_p \sigma_{\mu\nu} u_n~\overline{u_{N}} \sigma^{\mu\nu} P_L u_{\ell^-}^C \;.
\end{eqnarray}
In the main part of the text we translate the bounds in Secs.~4.4 and 4.5 of Ref.~\cite{Gonzalez-Alonso:2018omy} to our operator basis. The first case without RH neutrinos, which is discussed in Sec.~4.4, assumes for the Lee-Yang effective couplings~\cite{Lee:1956qn} $C_i'=C_i$, i.e. there are no couplings to right-handed neutrinos. The second fit, which is presented in Sec.~4.5, involves LH neutrinos for the vector and axial-vector interactions and RH neutrinos for the scalar and tensor interactions, i.e. $C'_V=C_V$, $C'_A=C_A$, $C'_S=-C_S$, $C'_T=-C_T$. The LNEFT Wilson coefficients introduced in Sec.~\ref{sec:LEFT} are related to the Lee-Yang effective couplings as follows
\begin{itemize}
\item LNC $\nu$ case:
\begin{eqnarray}\nonumber
&&{1\over 4}\Big(C_{ud\ell\nu 1}^{V}+C_{ud\ell\nu 2}^{V}\Big)g_V=C_V=C_V'\;,
{1\over 4}\Big(C_{ud\ell\nu 1}^{V}-C_{ud\ell\nu 2}^{V}\Big)g_A=-C_A=-C_A'\;,\\\nonumber
&&{1\over 4}\Big(C_{ud\ell\nu 1}^{S}+C_{ud\ell\nu 2}^{S}\Big)g_S=C_S=C_S'\;, ~
{1\over 4}\Big(C_{ud\ell\nu 1}^{S}-C_{ud\ell\nu 2}^{S}\Big)g_P=C_P=C_P'\;,\\
&&{1\over 2}C_{ud\ell\nu}^{T}g_T={1\over 2}C_T={1\over 2}C_T'\;,
\end{eqnarray}
\item LNC $N$ case:
\begin{eqnarray}\nonumber
&&{1\over 4}\Big(C_{ud\ell N 1}^{V}+C_{ud\ell N 2}^{V}\Big)g_V=C_V=-C_V'\;,
{1\over 4}\Big(C_{ud\ell N 1}^{V}-C_{ud\ell N 2}^{V}\Big)g_A=C_A=-C_A'\;,\\\nonumber
&&{1\over 4}\Big(C_{ud\ell N 1}^{S}+C_{ud\ell N 2}^{S}\Big)g_S=C_S=-C_S'\;,~
{1\over 4}\Big(C_{ud\ell N 1}^{S}-C_{ud\ell N 2}^{S}\Big)g_P=C_P=-C_P'\;,\\
&&{1\over 2}C_{ud\ell N}^{T}g_T={1\over 2}C_T=-{1\over 2}C_T'\;,
\end{eqnarray}
\item LNV $\nu$ case:
\begin{eqnarray}\nonumber
&&{1\over 4}\Big(C_{du\ell\nu 1}^{V\ast}+C_{du\ell\nu 2}^{V\ast}\Big)g_V=-C_V=C_V'\;,
{1\over 4}\Big(C_{du\ell\nu 1}^{V\ast}-C_{du\ell\nu 2}^{V\ast}\Big)g_A=-C_A=C_A'\;,\\\nonumber
&&{1\over 4}\Big(C_{du\ell\nu 1}^{S\ast}+C_{du\ell\nu 2}^{S\ast}\Big)g_S=C_S=-C_S'\;,~
{1\over 4}\Big(C_{du\ell\nu 1}^{S\ast}-C_{du\ell\nu 2}^{S\ast}\Big)g_P=C_P=-C_P'\;,\\
&&{1\over 2}C_{du\ell\nu}^{T\ast}g_T=-{1\over 2}C_T={1\over 2}C_T'\;,
\end{eqnarray}
\item LNV $N$ case:
\begin{eqnarray}\nonumber
&&{1\over 4}\Big(C_{du\ell N 1}^{V\ast}+C_{du\ell N 2}^{V\ast}\Big)g_V=-C_V=-C_V'\;,
{1\over 4}\Big(C_{du\ell N 1}^{V\ast}-C_{du\ell N 2}^{V\ast}\Big)g_A=C_A=C_A'\;,\\\nonumber
&&{1\over 4}\Big(C_{du\ell N 1}^{S\ast}+C_{du\ell N 2}^{S\ast}\Big)g_S=C_S=C_S'\;,~~~~~~~
{1\over 4}\Big(C_{du\ell N 1}^{S\ast}-C_{du\ell N 2}^{S\ast}\Big)g_P=C_P=C_P'\;,\\
&&{1\over 2}C_{du\ell N}^{T\ast}g_T=-{1\over 2}C_T=-{1\over 2}C_T'\;.
\end{eqnarray}
\end{itemize}
Hence we find for the relevant quark-level Wilson coefficients used in
Ref.~\cite{Gonzalez-Alonso:2018omy} in the first fit
\begin{align}
\epsilon_R&={\overline{C}_V+\overline{C}'_V\over 4g_V}+{\overline{C}_A+\overline{C}'_A\over 4g_A}\;,
&
\epsilon_S&={\overline{C}_S+\overline{C}'_S\over 2g_S}\;,
	&
\epsilon_P&={\overline{C}_P+\overline{C}'_P\over 2g_P}\;,
&
\epsilon_T&={\overline{C}_T+\overline{C}'_T\over 8g_T}\;,
\end{align}
where the Lee-Yang couplings $C_i$ are redefined as $C_i=(G_F V_{ud}/\sqrt{2})\overline{C}_i$.
The relevant quark-level Wilson coefficients in the second fit are
\begin{align}
  \tilde{\epsilon}_S
  &={\overline{C}_S-\overline{C}'_S\over 2g_S}\;,
  &
\tilde{\epsilon}_T&={\overline{C}_T-\overline{C}'_T\over 8g_T}\;.
\end{align}

\bibliographystyle{JHEP}
\bibliography{refs}

\end{document}